\documentclass[fleqn,usenatbib]{mnras}

\usepackage{newtxtext,newtxmath}
\usepackage[T1]{fontenc}
\usepackage{threeparttable}
\usepackage{booktabs}
\usepackage{url}
\usepackage{hyperref}
\usepackage{longtable}
\usepackage{graphicx}
\usepackage{listing}
\usepackage[T1]{fontenc}
\usepackage{color}
\DeclareRobustCommand{\VAN}[3]{#2}
\let\VANthebibliography\thebibliography
\def\thebibliography{\DeclareRobustCommand{\VAN}[3]{##3}\VANthebibliography}

\usepackage{graphicx}	
\usepackage{amsmath}	
\usepackage{amssymb}	
\usepackage{lineno}

\title{Properties of Gamma-Ray Bursts Associated with Supernovae and Kilonovae}
\author[Q. M. Li et al.]{
Q. M. Li$^{1}$,
Z. B. Zhang$^{1}$\thanks{E-mail:z-b-zhang@163.com},
X. L. Han$^{2}$,
K. J. Zhang$^{1}$,
X. L. Xia$^{1}$
and
C. T. Hao$^{1}$
\\
$^{1}$Department of Physics, college of physics, Guizhou university, Guiyang, 550025, China\\
$^{2}$Department of Physics and Astronomy, Butler University, Indianapolis, IN 46208, USA}
\date{Accepted XXX. Received YYY; in original form ZZZ}

\pubyear{2023}

\begin{document}
\label{firstpage}
\pagerange{\pageref{firstpage}--\pageref{lastpage}}
\maketitle

\begin{abstract}

We systematically compare the temporal and spectral properties of 53 Supernova (SN)-associated and 15 Kilonova (KN)-associated Gamma-Ray Bursts (GRBs). We find that the spectral parameters of both types GRBs are identically and lognormally distributed, consistent with those normal GRBs. The bolometric luminosities of SN/GRBs and KN/GRBs have a triple form with the corresponding break luminosities of SN/GRBs are roughly two orders of magnitude larger than those of KN/GRBs. We build the power-law relations between the spectral lag and the luminosity of prompt $\gamma$-rays with indices of $-1.43\pm0.33$ for SN/GRBs and $-2.17\pm0.57$ for KN/GRBs in the laboratory frame, which are respectively coincident with the rest-frame values. We verify that both SN/GRBs and KN/GRBs comply with their own Amati relations that match those of long and short GRBs, respectively. Analyzing X-ray afterglows with good plateau segments, we build the power-law relations between the X-ray luminosity and the plateau time with an index of $-1.12\pm0.17$ for KN/GRBs and $-1.08\pm0.22$ for SN/GRBs, which can be well explained by the relativistic shock driven by an energy injection. The plots of luminosity-lag, Amati relation and luminosity-time show heavy overlap between the two types of GRBs, implying that they might share the same radiation mechanism despite originating from different progenitors or central engines.
\end{abstract}




\section{Introduction}
Gamma-Ray Bursts (GRBs) are dramatic and transient events that emit high-energy electromagnetic radiations from various sources in the universe. These events are typically categorized as long or short GRBs with a dividing line at a duration time of $T_{90}\simeq2$ seconds \citep{1993ApJ...413L.101K,Zhang+08}. However, there is ongoing debate about the criteria used to classify GRBs based on $T_{90}$. For example, the observations from Burst and Transient Source Experiment (BATSE), \textit{Swift} Burst Alert Telescope (BAT) and \textit{Fermi} Gamma-ray Burst Monitor (GBM) support a bimodal $T_{90}$ classification \citep[e.g.,][]{1993ApJ...413L.101K,Zhang+08,Zhang+16,Zitouni-15,Tarnopolski+17,Zitouni-18}, although the BAT detector tends to detect softer $\gamma$-rays than the other two monitors. On the other hand, Some author argued that there are three \citep{Chattopadhyay+07,Horvath+16} or more subgroups of GRBs \citep{Chattopadhyay+18,Toth+19}. Recent investigations suggest that these additional components may be artificial \citep{Tarnopolski+19a,Tarnopolski+19b,2022ApJ...940....5D}.

Long GRBs (LGRBs) are thought to originate from the death of massive stars because they are associated with Supernovae (SNe, \citealt{Woosley2006,2012grb..book..169H}) and localized in star-forming galaxies \citep{2008ApJ...677L..85M}. The association between GRB 980425 and SN 1998bw was first identified by comparing the images of GRB 980425's host galaxy before and after appearance \citep{1998Natur.395..670G}. GRB 030329 was later confirmed to associate with a SN through spectroscopic analysis of its bright optical afterglows in a few days after the GRB. The SN spectrum could be separate after the spectral contributions of the GRB afterglow are subtracted \citep{2003ApJ...599..394M,2003ApJ...591L..17S,2004ApJ...609..952Z}. However, it's important to note that the duration of SN-associated GRBs could be shorter than the typical boundary of $\sim$2 s. It can be understood that the dividing line between the collapsing and non-collapsing GRBs is believed to be at $\sim$1 s \citep{Bromberg2013,2022ApJ...940....5D} which is close to the dividing line derived for Swift GRBs \citep{Zhang2020}. For example, short GRB 200826A has a duration of ${T_{90}}\approx $1.14s \citep{2021NatAs...5..917A,2020GCN.28949....1R}, but it is likely associated with a SN resembling SN 1998bw and could be alternatively accompanied by a Kilonova (KN) with larger uncertainties \citep{2022ApJ...932....1R}. Meanwhile, \cite{2021NatAs...5..911Z} pointed out that the observed properties of GRB 200826A such as spectral behaviours, total energy and host galaxy are similar to those of long GRBs \citep{2021NatAs...5..911Z}.

Short GRBs (SGRBs) are commonly believed to originate from the binary mergers of compact objects involving either double neutron stars (DNS) or neutron star- black hole (NS-BH) systems \citep{1984SvAL...10..177B,1986ApJ...308L..43P}. The merger of compact objects produces relativistic jet that generates the instantaneous $\gamma$-ray emissions together with multi-wavelength afterglows \citep{2011ApJ...732L...6R,2015ApJ...806L..14P,2016ApJ...824L...6R}. Some SGRBs show an additional component above the afterglows. For instance, GRB 130603B showed a bump light curve in the near-infrared band. This excess emission is interpreted as a kilonova resulting from the collision of two compact stars \citep{2016AdAst2016E...8T,2013ApJ...778L..16H,2013Natur.500..547T}. The first Gravitation-Wave-associated SGRB 170817A was detected by \textit{Fermi} Gamma-ray Burst Monitor and verified to result from the merge of a DNS system accompanied by an important optical transient AT 2017gfo \citep{2020MNRAS.493.3379R,2021ApJ...916...89R}, which provides a direct observation of kilonova \citep{2017ApJ...848L..14G,2017Abbott}. Additionally, an obvious emission excess in the near-infrared band, consistent with a KN being powered by a r-process, was also discovered \citep{2017Natur.551...80K,2020arXiv201204810M}. The KN rate of AT 2017gfo-like events was constrained to be less than 1775\ Gpc$^{-3}$yr$^{-1}$ using a survey simulation model \citep{2020ApJ...904..155A}. It is notable that the multi-wavelength kilonova components of a SGRB usually peak around several days after the burst, which is significantly earlier than the peaking times of a supernova connected with the corresponding LGRB \cite[see e.g.][]{2015ApJ...811L..22J}. Moreover, the KN-associated GRBs, like GRB 211211A, can originate from the binary mergers but have persistent prompt $\gamma$-ray emissions lasting longer than 2 s \citep{2022Natur.612..223R,2022Natur.612..232Y,2022ApJ...931L..23L}. This demonstrates that the $T_{90}$ should be combined with other parameters to constrain the true origins of GRBs reasonably \cite[see also e.g.][]{2009ApJ...703.1696Z,2021NatAs...5..911Z}.

Using a set of seven long GRBs observed by CGRO/BATSE and BeppoSAX, \cite{2000ApJ...534..248N} proposed the first anti-correlation between peak luminosity ($L_p$) and spectral lag ($\tau$) of six normal bursts with the exception of SN/GRB 980425 as $L_p\propto\tau^{-1.15}$. It is obviously seen that GRB 980425/SN1998bw is an outlier of this relation. The anti-correlation was suggested to be influenced by the initial bulk Lorentz factor ${\Gamma _0}$ of GRB outflows \citep{2000ApJ...544L.115S,2001ApJ...563L.123S,2003MNRAS.342..587D,2006MNRAS.373..729Z}. In addition, some KN-associated GRBs have been found to violate the power-law relation of $L_p$ with $\tau$. GRB 060614 with a very long duration of $T_{90}\simeq108.7$ s however behaves the typical features of a SGRB such as the extended emissions after main peak and the negligible spectral lag. It was discovered that GRB 060614 is associated with a KN on basis of the observations of its multi-band afterglows \citep{2015NatCo...6.7323Y,2015ApJ...811L..22J}. Late on, \cite{2008ApJ...677L..85M} reported that this burst was clearly inconsistent with the lag-luminosity relation discovered by \cite{2000ApJ...534..248N} for six LGRBs. They also noticed that two more low-luminosity SN/GRBs 060218 and 031203 were similarly distributed away from the $\tau- L_p$ relation found by \cite{2000ApJ...534..248N}. This motivates us to investigate whether the SN- and KN-associated GRBs have their own lag-luminosity relations and identify their differences.

Based on the current multi-band observations of kilonovae and supernovae, we aim to systematically investigate their similarities and differences in the temporal and spectral aspects of both $\gamma$-rays and X-rays systematically in this paper. In Section 2, we describe our sample selection criteria and provide details on the data preparation process. Section 3 presents our data reduction method and main results, which will be presented in a clear and concise manner. Finally, in Section 4, we summarize our findings and provide conclusions based on our analyses.

\section{Data preparation}
 \label{sec:sample}
In order to systematically compare the temporal and spectral properties of SN- and KN-associated GRBs in the X- and $\gamma$-ray bands, we have compiled a sample of 68 GRBs, we firstly collect 68 GRBs including 53 SN/GRBs and 15 KN/GRBs from literature published between 1997 and 2020 \citep[e.g.][]{2021AAS...23723306D}. Our sample selection criteria are as follows: (1) GRBs should consist of significant pulse structures in channels 1 (15-25 keV) and 3 (50-100 keV) so that the time lags can be accurately measured; (2) There should be enough data points of X-ray afterglow curves in order to identify their temporal profiles; (3) The redshift of each GRB in our samples should be available. Tables \ref{tab:1} and \ref{tab:2} present the primary parameters of the SN- and KN-associated GRBs, respectively, in which the broken power-law (or Band function) spectrum of GRBs has been assumed \citep{Band1993,1995ApJ...439..307F,2008ApJ...679..570S} since the spectral form can be adopted to describe almost all GRBs \citep{Zhang2020}. The tables list three typical parameters: low/high-energy indices ($\alpha/\beta$) and peak energy $\textit{E}_{\rm p}$. Interestingly, the BAND function will degenerate into the cut-off power-law (CPL) form as the $\beta$ equals to negative infinity. Note that all GRBs in our samples were detected by the satellites such as \textit{Swift}/BAT, \textit{Swift}/XRT, CGRO/BATSE, HETE-II, \textit{BeppoSAX}, Konus-\textit{wind} or \textit{Fermi}/GBM.

In Tables \ref{tab:1} and \ref{tab:2}, we list some of the typical parameters of our GRB samples. Columns 1-2 show the GRB names together with their detecting satellites. Column 3 gives the burst duration ${T_{90}}$ in $\gamma$-rays. The redshift is given in column 4. Columns 5-6 respectively represent the 1s peak photon flux $P$ and fluence ${S_\gamma}$. Columns 7-9 provide three representatively spectral parameters. Column 10 displays the energy range of detectors. Column 11 is the spectral index $\beta_x$ of X-rays. The relevant references are listed in Column 12. It is worth noting that most SN/GRBs except GRB 200826A are long-duration bursts with $T_{90}$$>$2 s, while 11 out of 15 KN/GRBs are short-duration GRBs. Meanwhile, 8 out of 53 SN/GRBs are superluminous SN-associated (SLSN) GRBs in our sample. However, it is important to mention that GRB 060707 lacks spectroscopic evidence of the SN component. GRBs 090424, 100621A and 150821A exhibit bumps that are inconsistent with the SN/GRBs or have much lower significance. Therefore,we have excluded these four bursts from our analysis. While our sample of SN/KN-associated GRBs may not satisfy completeness conditions absolutely, they are sufficient for statistical study. With the increasing numbers of two classes of bursts, the complete samples could be more helpful for their comparisons.

\section{Methods and results}
\label{Data Analysis Method}
\subsection{Comparison of SN/KN-associated GRBs in prompt $\gamma$-rays}
\subsubsection{Parametric distributions}
\begin{figure}
	\includegraphics[width=\columnwidth]{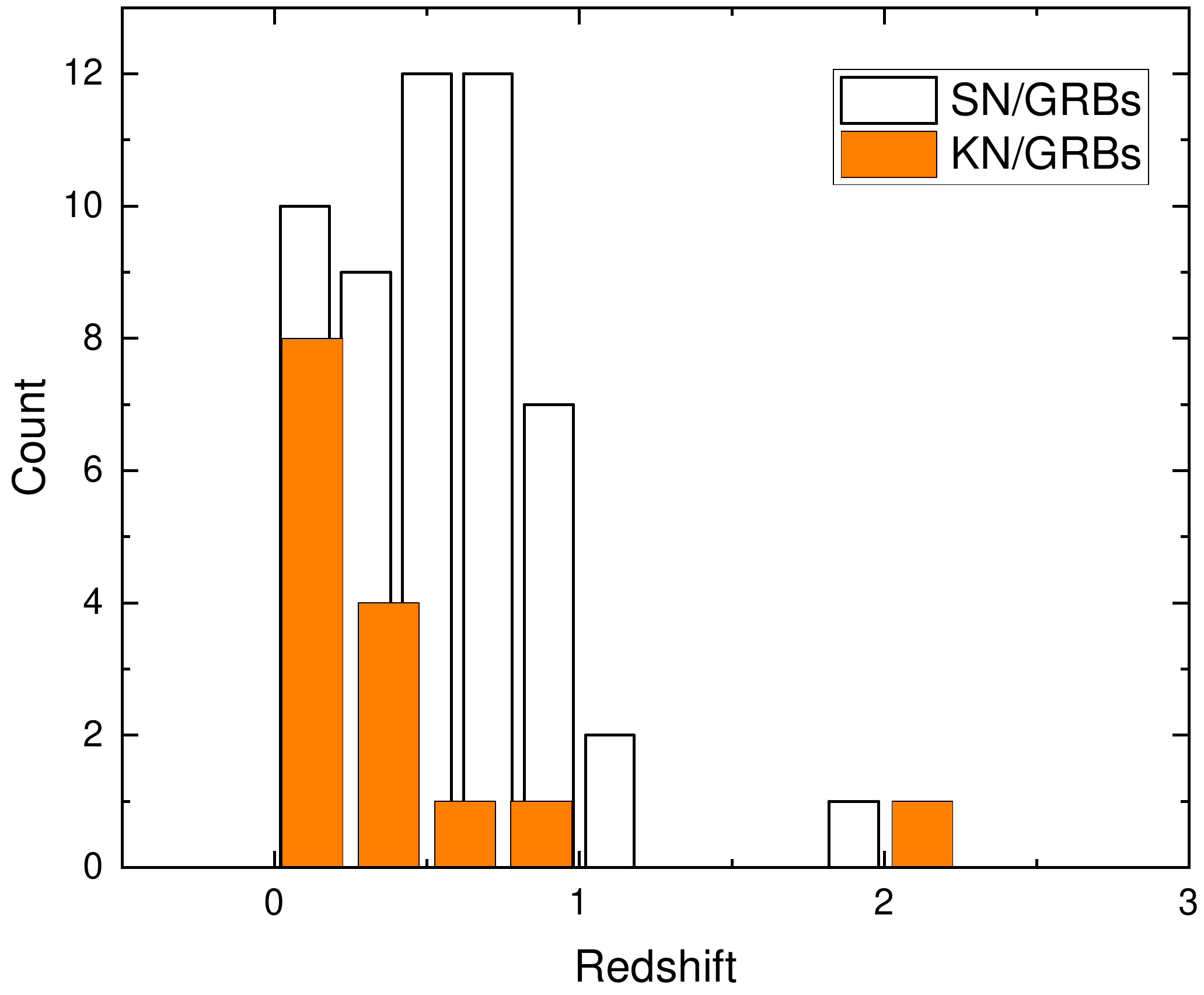}
   \caption{The histograms of redshift distributions of 53 SN/GRBs (empty) and 15 KN/GRBs (filled).}
	\label{fig1:redshift}
\end{figure}

\begin{figure*}
	\centering
	\includegraphics[width=\columnwidth]{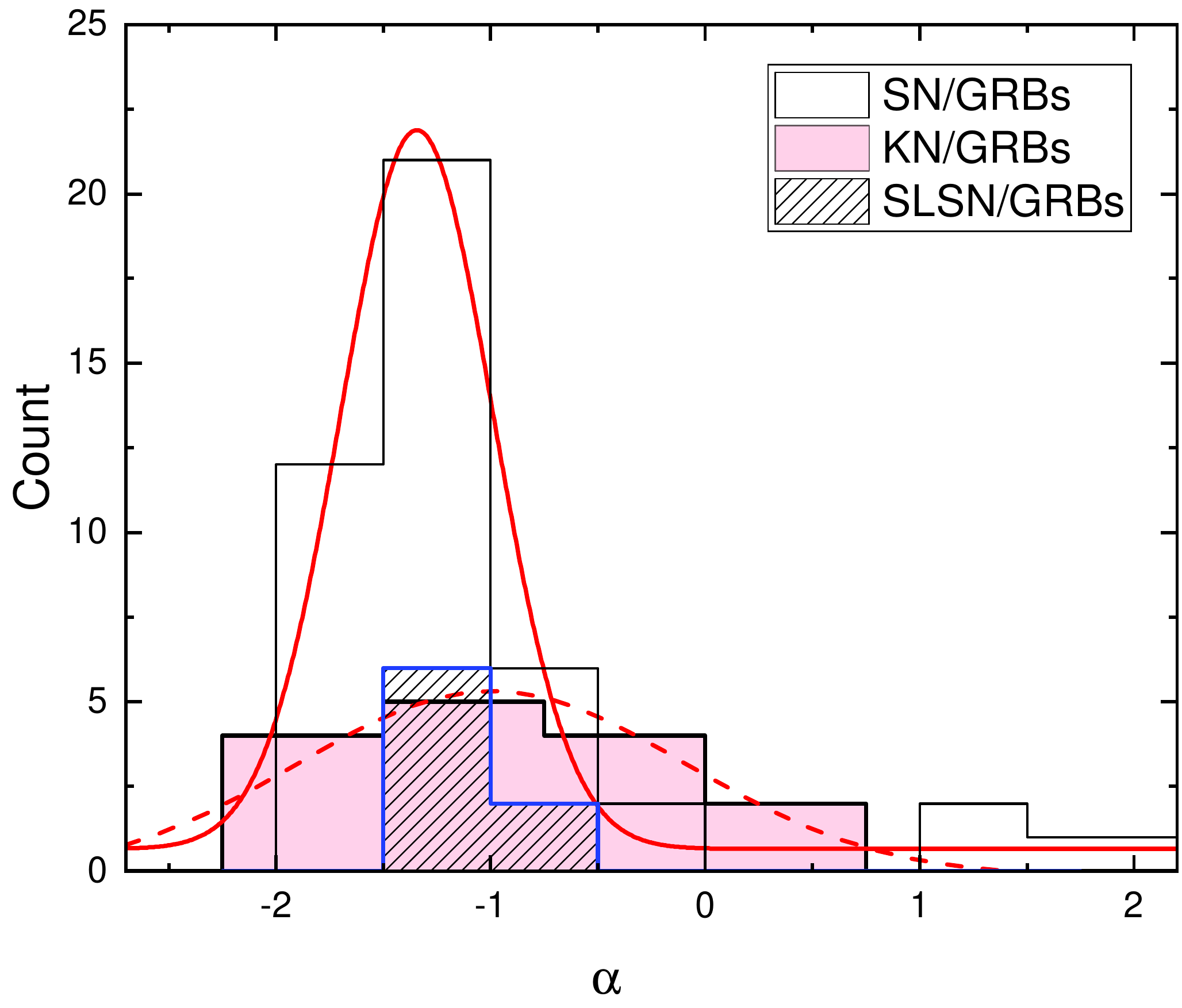}
\includegraphics[width=\columnwidth]{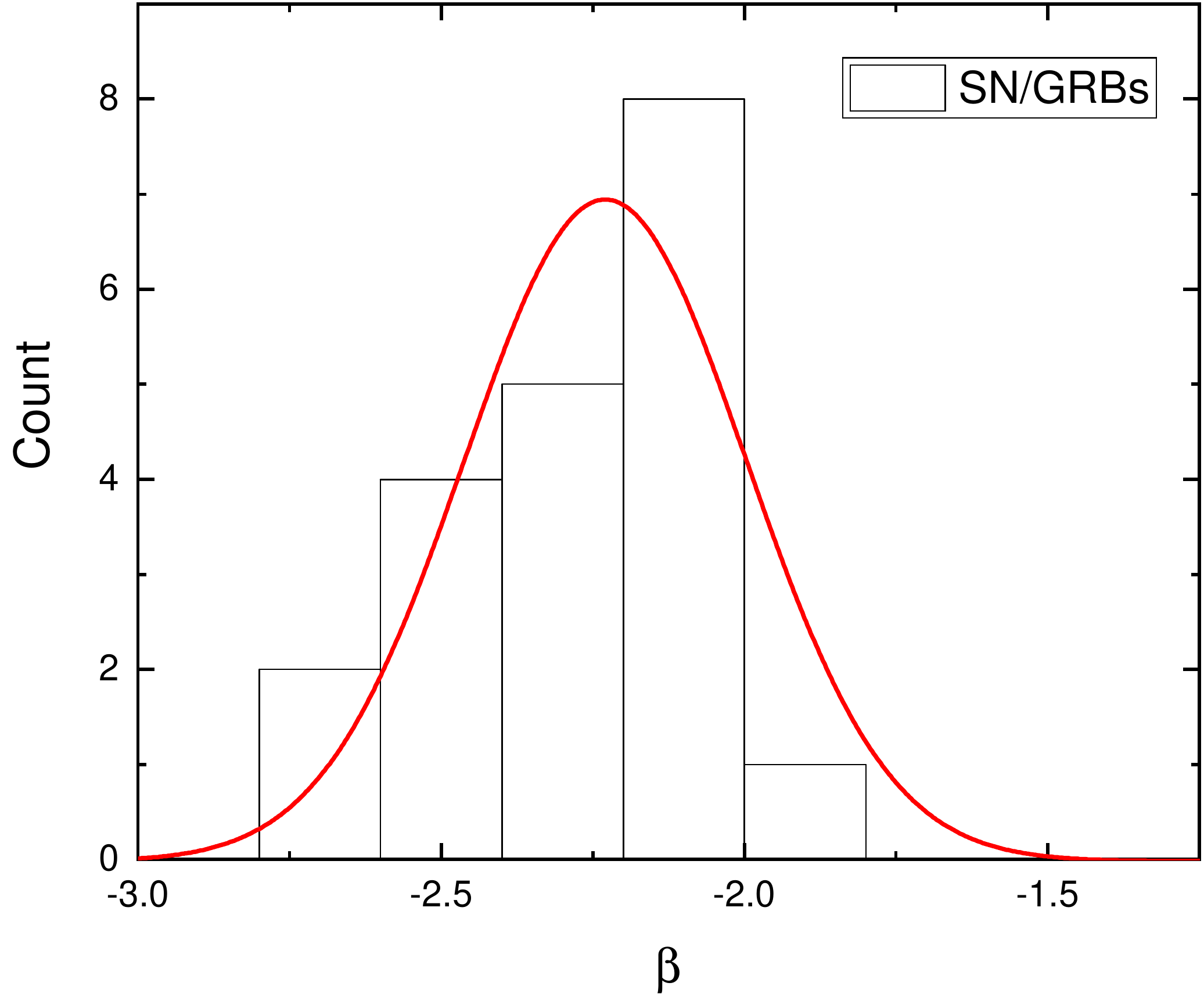}
\includegraphics[width=\columnwidth]{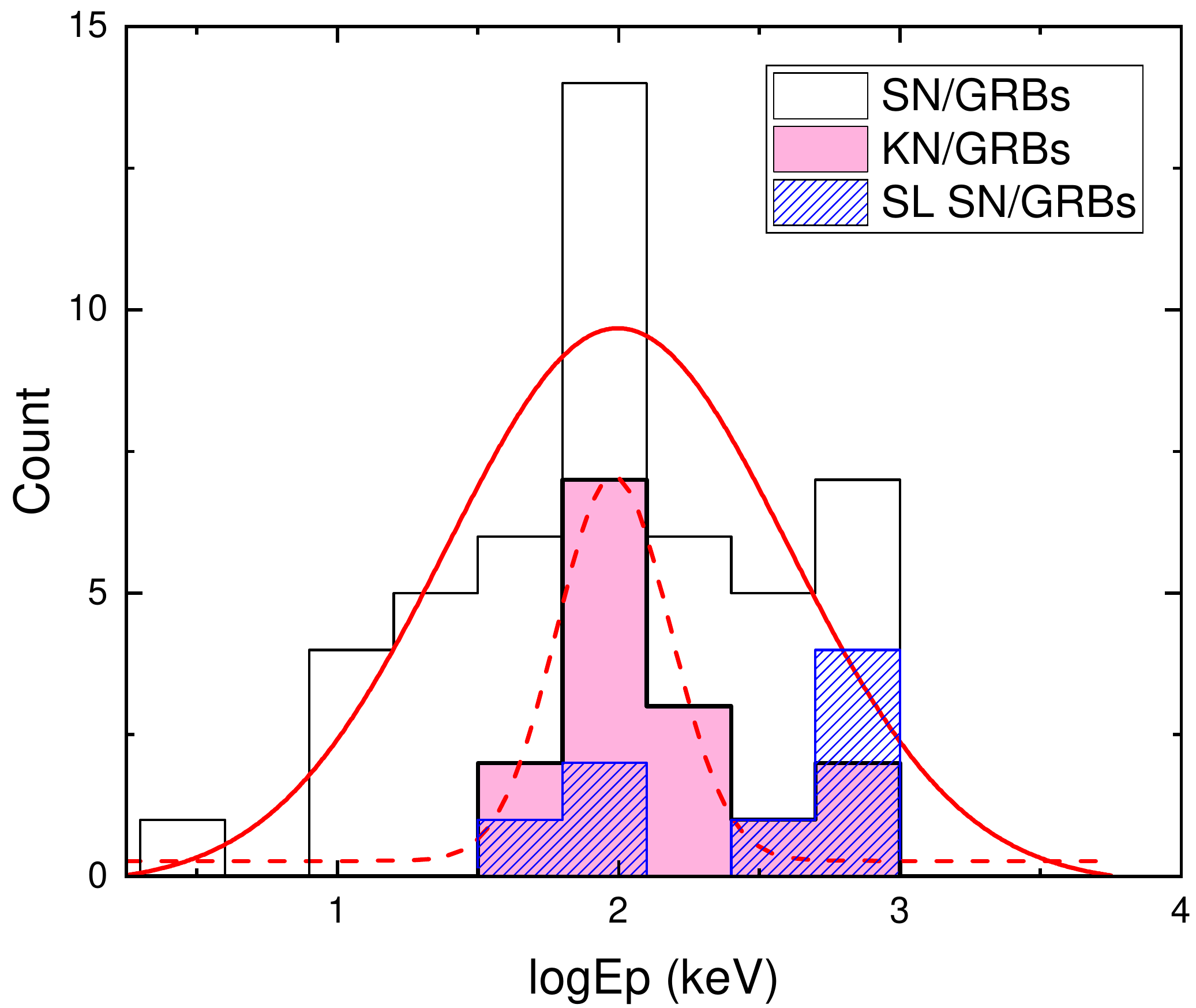}
	\caption{The distributional histograms of low- and high-energy photon indexes ($\alpha$ and $\beta$) and peak energy $E_p$ for the samples of SN/GRBs (blank), KN/GRBs (filled) and SLSN/GRBs (hatched). The solid and the dash lines denote the best fits to them with a Gauss function.}
	\label{fig2:alpha-Ep}
\end{figure*}

The redshift distributions of SN- and KN-associated GRBs are displayed in Figure \ref{fig1:redshift}, where we find that both of them are located at the nearby universe and the median redshift of SN/GRBs is about 0.5 that is roughly two times larger than that of KN/GRBs. Figure \ref{fig2:alpha-Ep} shows the $\alpha$ and $E_p$ distributions of the SN/KN-associated and SLSN GRBs. The mean value of $\alpha$ to be -1.34$\pm$0.02 with a standard error of 0.35 for SN/GRBs and -1.09$\pm$0.15 with a standard error of 1.01 for KN/GRBs, which is close to that of SLSN/GRBs. The mean value of high-energy spectral index of SN/GRBs is $\langle\beta\rangle\approx-2.23\pm0.03$ with a scatter of 0.23. The averaged peak energies are respectively $97.7^{+28.2}_{-21.8}$ keV and $95.5^{+7.0}_{-5.9}$ keV for the SN/GRBs and the KN/GRBs, which are comparable to those of \textit{Swift} GRBs \citep{2018PASP..130e4202Z} but significantly less than the typical $E_p$ values of $\sim200$ keV for the long and short GRBs measured by wider detectors \citep{2000ApJS..126...19P,Preece2016}. We note that the narrower energy band of \textit{Swift}/BAT does not affect our $E_p$ distributions since the larger $E_p$ values of \textit{Swift} GRBs with hard spectra peaking out of the BAT range have been measured by the joint spectra with other satellites. This demonstrates that both SN/GRBs and KN/GRBs have softer spectra than those normal GRBs. The observed peak energies of SN/GRBs span about two orders of magnitude, which is wider than the $E_p$ range of KN/GRBs. However, two spectral power-law indexes ($\alpha$ and $\beta$) are consistent with previous values of normal bursts \cite[see e.g.][]{2000ApJS..126...19P,2018PASP..130e4202Z}, indicating that the dominant radiation mechanisms of both SN/GRBs and KN/GRBs are synchronous acceleration of electrons.

\begin{figure*}
	\centering
	\includegraphics[width=\columnwidth]{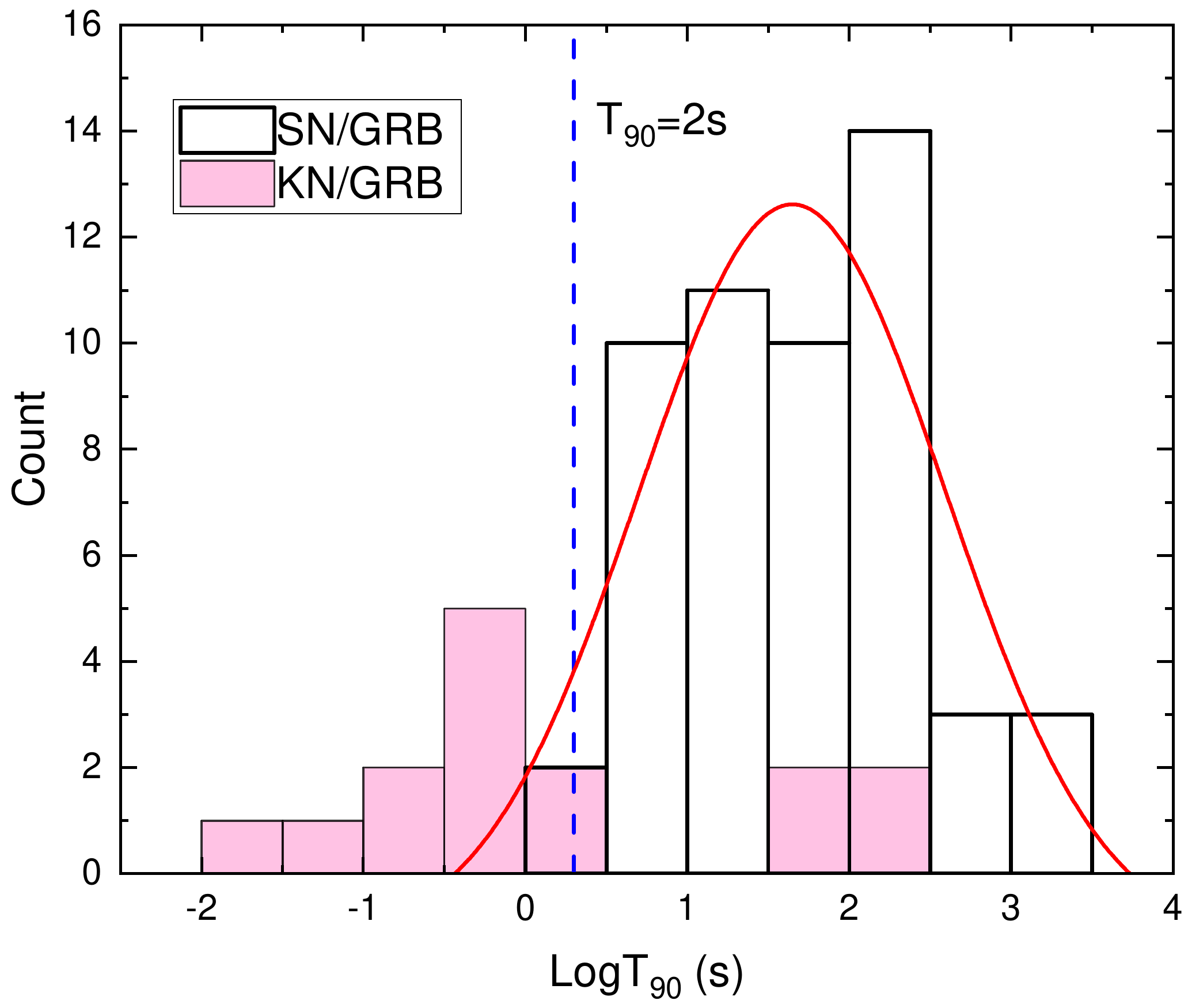}
    \includegraphics[width=\columnwidth]{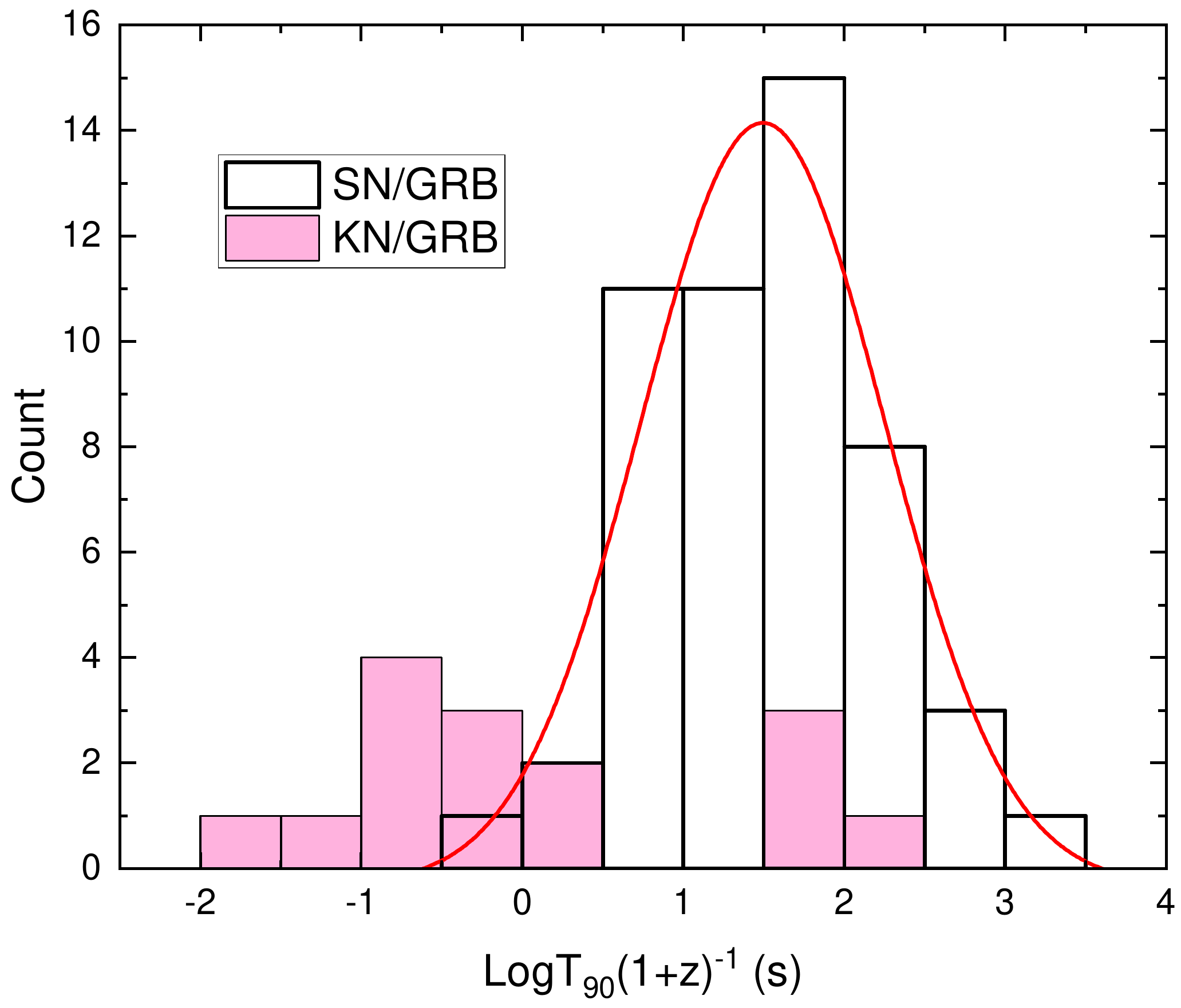}
   	\caption{The lognormal distributions of time duration of SN/KN-associated GRBs in the observer frame (left panel) and in the rest frame (right panel). The red solid lines stand for the best fit to SN/GRBs with a Gaussian function.}
	\label{fig:T90dis}
\end{figure*}

Since KN/GRBs and SN/GRBs are usually thought to result from the mergers of compact binaries and the core-collapse of massive stars, respectively, of which both of them should exhibit distinct time durations in prompt $\gamma$-rays. Hence, we plot the distributions of the observed time duration $T_{90}$ and the intrinsic one $T_{90,i}\equiv T_{90}(1+z)^{-1}$ in Figure \ref{fig:T90dis}, where we confirm that KN/GRBs do have shorter durations than SN/GRBs as a whole, which is similar to the distributional features of \textit{Swift}/BAT bursts \citep{Zhang+08}. The $T_{90}$ values of SN/GRBs are log-normally distributed with a mean of $44.67^{+16.98}_{-12.31}$ s and a scatter of 1.90 dex, which is good in agreement with $42.66^{+2.00}_{-1.92}$ s gotten by \cite{Zhang2020} for the \textit{Swift} long GRBs with well-measured spectra. Similarly, the $T_{90,i}$ distribution seems to be bimodal \citep[see also][]{Zhang+08} and the SN/GRBs peaks at $30.9^{+6.25}_{-5.20}$ s with a scatter of 1.52 dex in the rest frame. It is worth noting that the numerary ratio of KN/GRBs to SN/GRBs is 15/53 which is much higher than $\sim 1/10$ for \textit{Swift}/BAT but very close to $\sim1/3$ of \textit{Fermi}/GBM or CGRO/BATSE bursts \citep{Zhang2020,2022ApJ...940....5D}.

\subsubsection{Multi-component light-curves of prompt $\gamma$-rays}

Here, we apply the signal-to-noise (S/N) method and the criteria proposed by \cite{2021ApJS..252...16L} to identify the different $\gamma$-ray components involving precursors, main peaks and extended emissions (EEs), in both SN/GRBs and KN/GRBs. To ensure a higher confidence level and avoid the selection effect of energy bands, we set a standard of S/N$\geq$3 in our data processing of \textit{Swift} GRBs only. In total, we have selected 30 SN/GRBs and 13 KN/GRBs detected by the \textit{Swift}/BAT with well-measured prompt $\gamma$-rays to constitute our gold samples. The results are given in Tables \ref{tab:3} and \ref{tab:4}, from which we can find that most SN/GRBs are untriggered for precursors and have multiple radiation components in their prompt $\gamma$-ray phases. The fraction of SN/GRBs with precursors is as high as 22/30$\sim$73\%. On the contrary, the majority of KN/GRBs are triggered for precursors and the fraction of KN/GRBs with precursors is about 6/13$\sim$46.2\%. Furthermore, the comparable fractions of bursts with EE are 56.7\% and 53.3\% for the \textit{Swift} SN/GRB and KN/GRB samples, respectively.
\subsubsection{Luminosity distribution}
\begin{figure}
	\centering
    \includegraphics[width=\columnwidth]{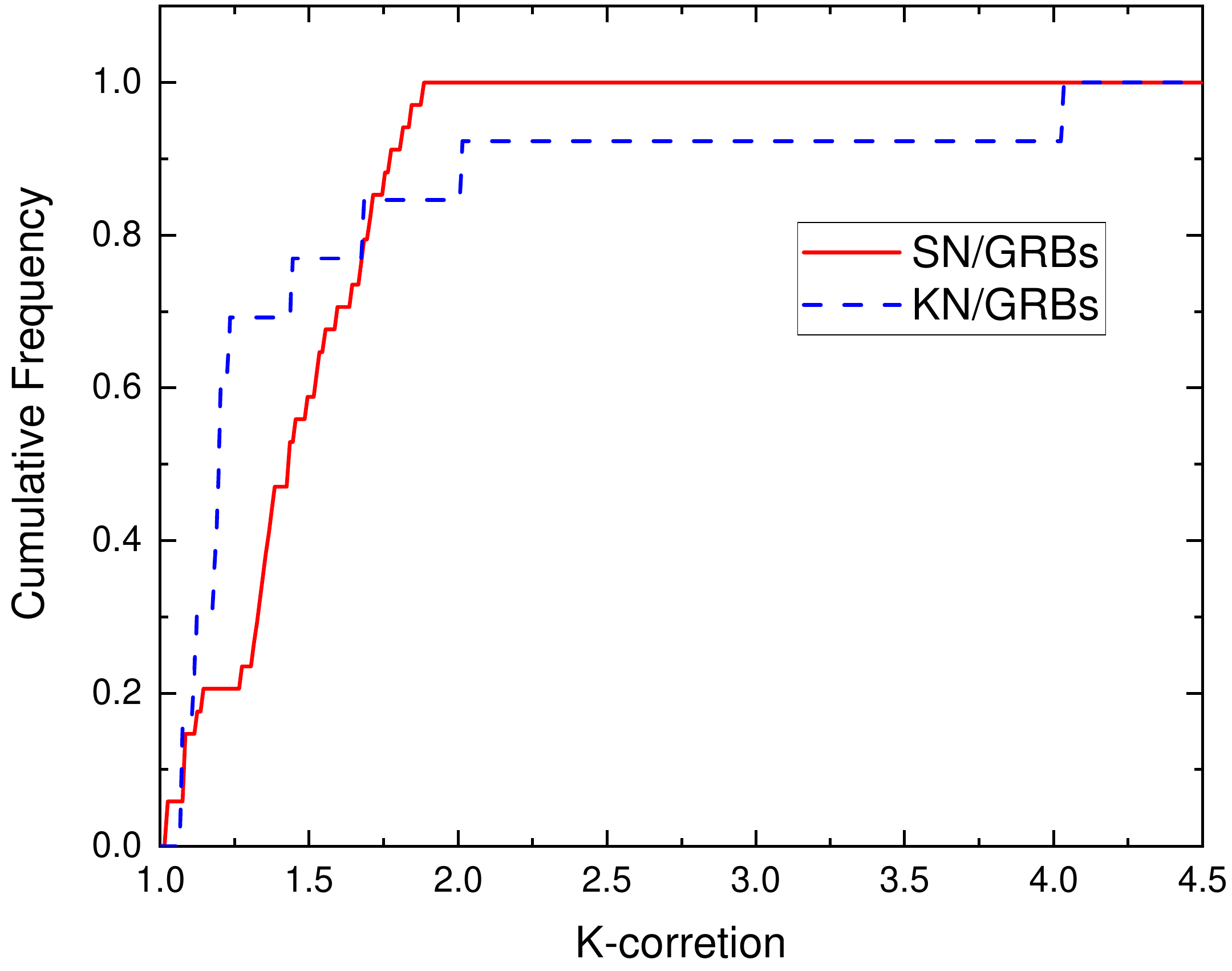}
	\caption{Cumulative fractions of k-correction for SN/GRBs (red solid line) and KN/GRBs (blue dash line).}
	\label{fig:kc}
\end{figure}
\begin{figure*}
	\centering
	\includegraphics[width=\columnwidth]{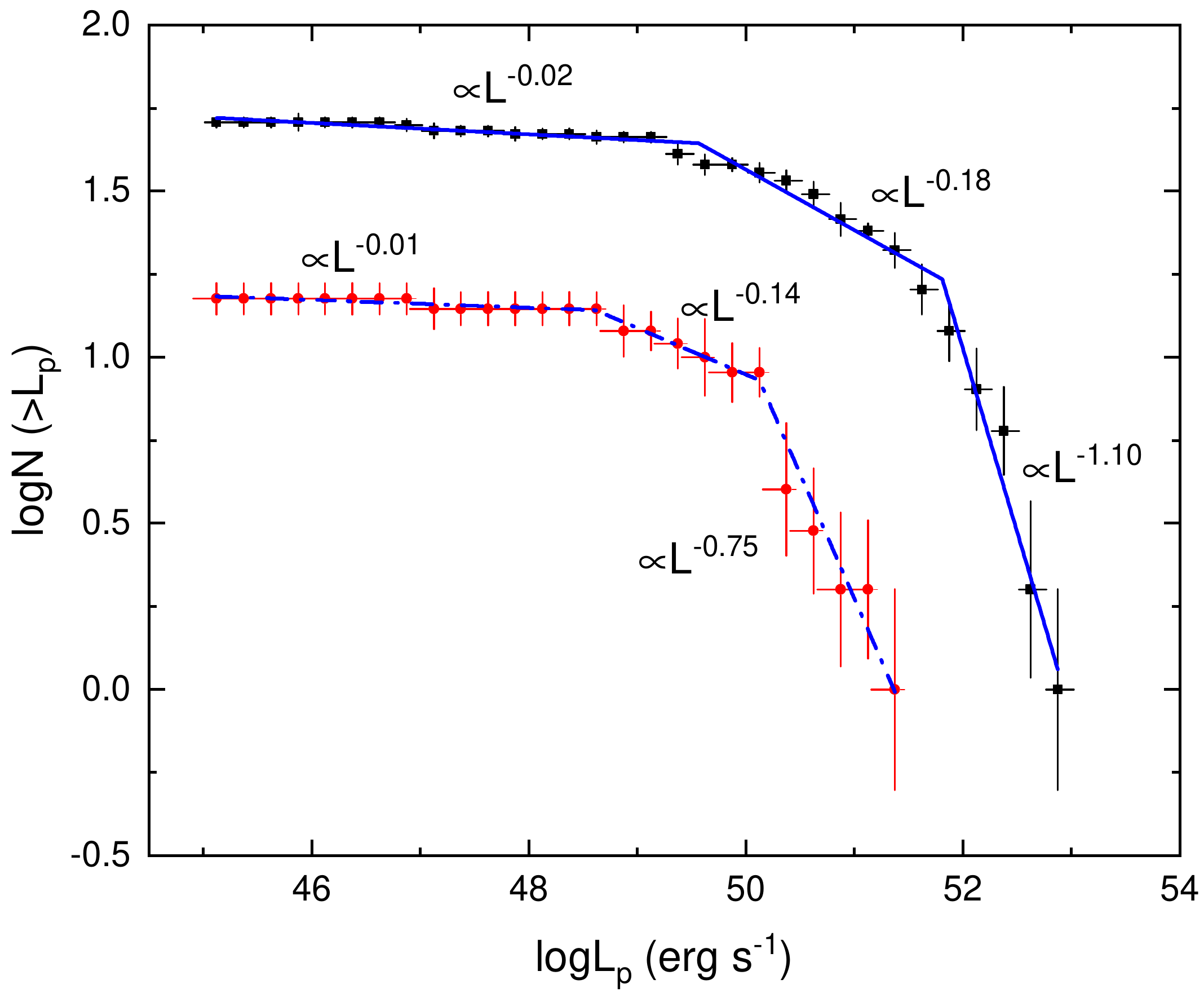}
\includegraphics[width=\columnwidth]{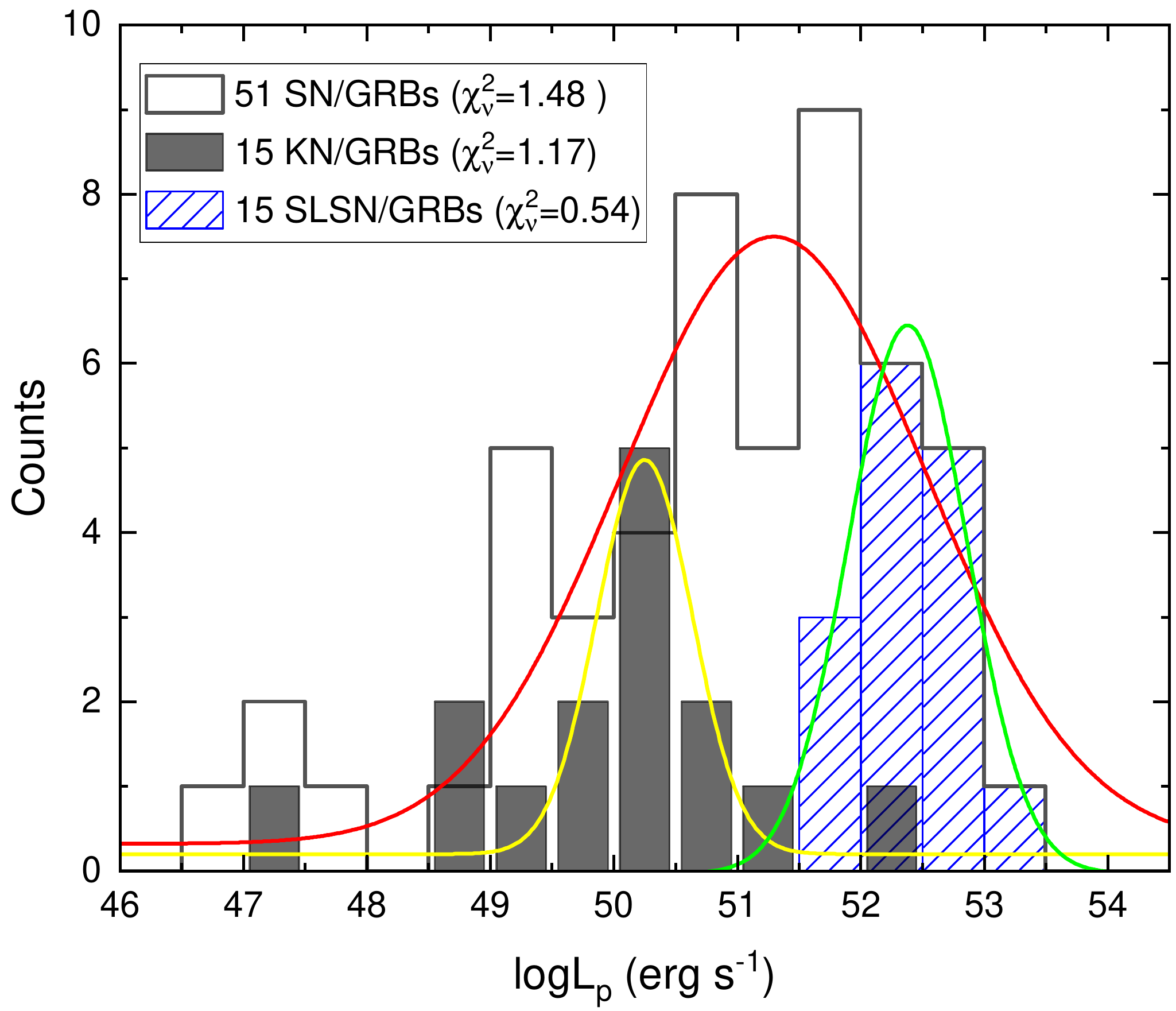}
	\caption{Left panel: cumulative luminosity distributions of 51 SN/GRBs (filled squares) in \textbf{Table \ref{tab:5}} and 15 KN/GRBs (filled circles) in Table \ref{tab:6}. The solid and the dash-dotted curves are the best fits with Equations (\ref{equ:2}) and (\ref{equ:4}) to the SN/GRB and KN/GRB data, respectively. Right panel: histograms of luminosities of three kinds of GRBs are well fitted with a lognormal function.}
	\label{fig:lpdis}
\end{figure*}
To investigate the potential application of SN/KN-associated GRBs in cosmology, we now follow our previous work \citep{2018PASP..130e4202Z} to calculate the peak luminosity $L\equiv L_p=4\pi D_{L}^{2}(z)PK_c$. Figure \ref{fig:kc} displays the cumulative distributions of the k-corrections ($K_c$) for two kinds of bursts, of which the median values are about $\sim$1.25 for the KN/GRBs and $\sim$1.5 for the SN/GRBs. Then we comparably study the cumulatively bolometric luminosity distributions of SN/GRBs and KN/GRBs in Figure \ref{fig:lpdis}, from which we fit the SN/GRB distribution of peak luminosities with a triple power-law form as follows
\begin{equation}
\label{sn:lpdis}
\log N_{SN}( > L) = \left\{{\begin{array}{*{20}{c}}
{\log {N_{b1}} + {{\rm{k}}_1}\log \left( {{L \mathord{\left/
 {\vphantom {L {{L_{b1}}}}} \right.
 \kern-\nulldelimiterspace} {{L_{b1}}}}} \right)} \hfill [L\leq L_{b1}],\\
{\log {N_{b1}} + {{\rm{k}}_2}\log \left( {{L \mathord{\left/
 {\vphantom {L {{L_{b1}}}}} \right.
 \kern-\nulldelimiterspace} {{L_{b1}}}}} \right)} \ \ \ \hfill [L_{b1}< L\leq L_{b2}],\\
{\log {N_{b2}} + {{\rm{k}}_3}\log \left( {{L \mathord{\left/
 {\vphantom {L {{L_{b2}}}}} \right.
 \kern-\nulldelimiterspace} {{L_{b2}}}}} \right)} \hfill [L> L_{b2}],
\end{array}} \right.
\end{equation}
where $N_{b1}$, $N_{b2}$, $k_1$, $k_2$, $k_3$, $L_{b1}$ and $L_{b2}$ are seven fitting parameters of which log$N_{b2}$= log$N_{b1}$+$k_2$log($L_{b2}/L_{b1}$), $N_{b1}\simeq36.31^{+5.41}_{-0.81}$, $k_1=-0.02\pm0.003$, $k_2=-0.27\pm0.014$ and $k_3=-0.99\pm0.02$. $L_{b1}=1.12_{-0.17}^{+0.20}\times {10^{{\rm{50}}}}{\rm{erg}}/\rm s$ and $L_{b2}=6.46_{-0.43}^{+0.46}\times {10^{{\rm{51}}}}{\rm{erg}}/\rm s$ are two broken luminosities. The triple power-law relation described in Eq. (\ref{sn:lpdis}) can be roughly expressed to be
\begin{equation}
\label{equ:2}
\begin{aligned}
 N_{SN}( > L)\propto\left\{{\begin{array}{*{20}{c}}
{L^{k_1}\sim L^{-0.02\pm0.003}} \hfill [L\leq L_{b1}],\\
{L^{k_2}\sim L^{-0.27\pm0.014}}\ \ \ \hfill [L_{b1}< L\leq L_{b2}],\\
{L^{k_3}\sim L^{-0.99\pm0.02}}  \hfill [L> L_{b2}],
\end{array}} \right.
\end{aligned}
\end{equation}

Note that the SN/GRBs with higher luminosity of $L>L_{b2}$ exhibit faster decrease of luminosity distribution and are defined as the SLSN/GRBs that have been symbolized with stars in Figure \ref{fig:lag and lag/1+z}. Similarly, we also fit the luminosity distribution of KN/GRBs with a triple power-law form in Figure \ref{fig:lpdis} as

\begin{equation}
\label{equ:3}
\log N_{KN}( > L) = \left\{{\begin{array}{*{20}{c}}
{\log {N_{b3}} + {{\rm{k}}_4}\log \left( {{L \mathord{\left/
 {\vphantom {L {{L_{b3}}}}} \right.
 \kern-\nulldelimiterspace} {{L_{b3}}}}} \right)} \hfill [L\leq L_{b3}],\\
{\log {N_{b3}} + {{\rm{k}}_5}\log \left( {{L \mathord{\left/
 {\vphantom {L {{L_{b3}}}}} \right.
 \kern-\nulldelimiterspace} {{L_{b3}}}}} \right)} \ \ \ \hfill [L_{b3}< L\leq L_{b4}],\\
{\log {N_{b4}} + {{\rm{k}}_6}\log \left( {{L \mathord{\left/
 {\vphantom {L {{L_{b4}}}}} \right.
 \kern-\nulldelimiterspace} {{L_{b4}}}}} \right)} \hfill [L> L_{b4}],
\end{array}} \right.
\end{equation}

where $\log {N_{b4}} = \log {N_{b3}} + {k_4}\log ({L_{b4}}/{L_{b3}})$,$N_{b3}\simeq13.80^{+0.09}_{-0.47}$, $k_4=-0.01\pm0.003$, $k_5=-0.14\pm0.02$, $k_6=-0.75\pm0.04$, $L_{b3}=4.07^{+1.17}_{-0.91}\times10^{48} \rm erg/\rm s$ and $L_{b4}=1.35^{+0.16}_{-0.15}\times10^{49} \rm erg/\rm s$ are seven fitting parameters. The broken power-law form of Eq. (\ref{equ:3}) can be simply read as
\begin{equation}
\label{equ:4}
\begin{aligned}
 N_{KN}( > L)\propto\left\{{\begin{array}{*{20}{c}}
{L^{k_4}\sim L^{-0.01\pm0.003}} \hfill [L\leq L_{b3}],\\
{L^{k_5}\sim L^{-0.14\pm0.02}}\ \ \ \hfill [L_{b3}< L\leq L_{b4}],\\
{L^{k_6}\sim L^{-0.75\pm0.04}}  \hfill [L> L_{b4}],
\end{array}} \right.
\end{aligned}
\end{equation}
in which we need to point out that the steeper decline of KN/GRBs with higher luminosity is due to rapid decrease of event rates in comparison to the second decline of SN/GRBs. However, the first power-law decay of two types of GRBs match each other in a coincident way, indicating the similarity of their spatial distribution at the end of lower luminosity. Additionally, we also want to emphasize that the Malmquist effects \citep{1997ARA&A..35..101T,2018PASP..130e1001D} and the sample selection effects \citep{2022MNRAS.513.1078D} will somewhat bias our results more or less, which has been beyond the scope of this paper. Figure \ref{fig:lpdis} shows the histograms of different kinds of GRBs and a K-S test gives $D=0.44$ (${D_\alpha }=0.40$) and $p=0.02$ for 51 SN/GRBs and 15 KN/GRBs. For 51 SN/GRBs and 15 SLSN/GRBs, the K-S test results are $D=0.71$ (${D_\alpha }=0.40$) and $p=3.78 \times {10^{ - 6}}$. Similarly, the K-S test results of 15 SLSN/GRB and 15 KN/GRB are $D=0.93$ (${D_\alpha }=0.50$) and $p=3.86 \times {10^{ - 7}}$ \textbf{(see Table \ref{tab:7} for details)}. This demonstrates that the three subgroups of GRBs are taken from distinct parent distributions.

\subsubsection{The Lag-Luminosity relation}
\label{subsec: galaxy-targeted}
\begin{figure*}
	\centering
	\includegraphics[width=\columnwidth]{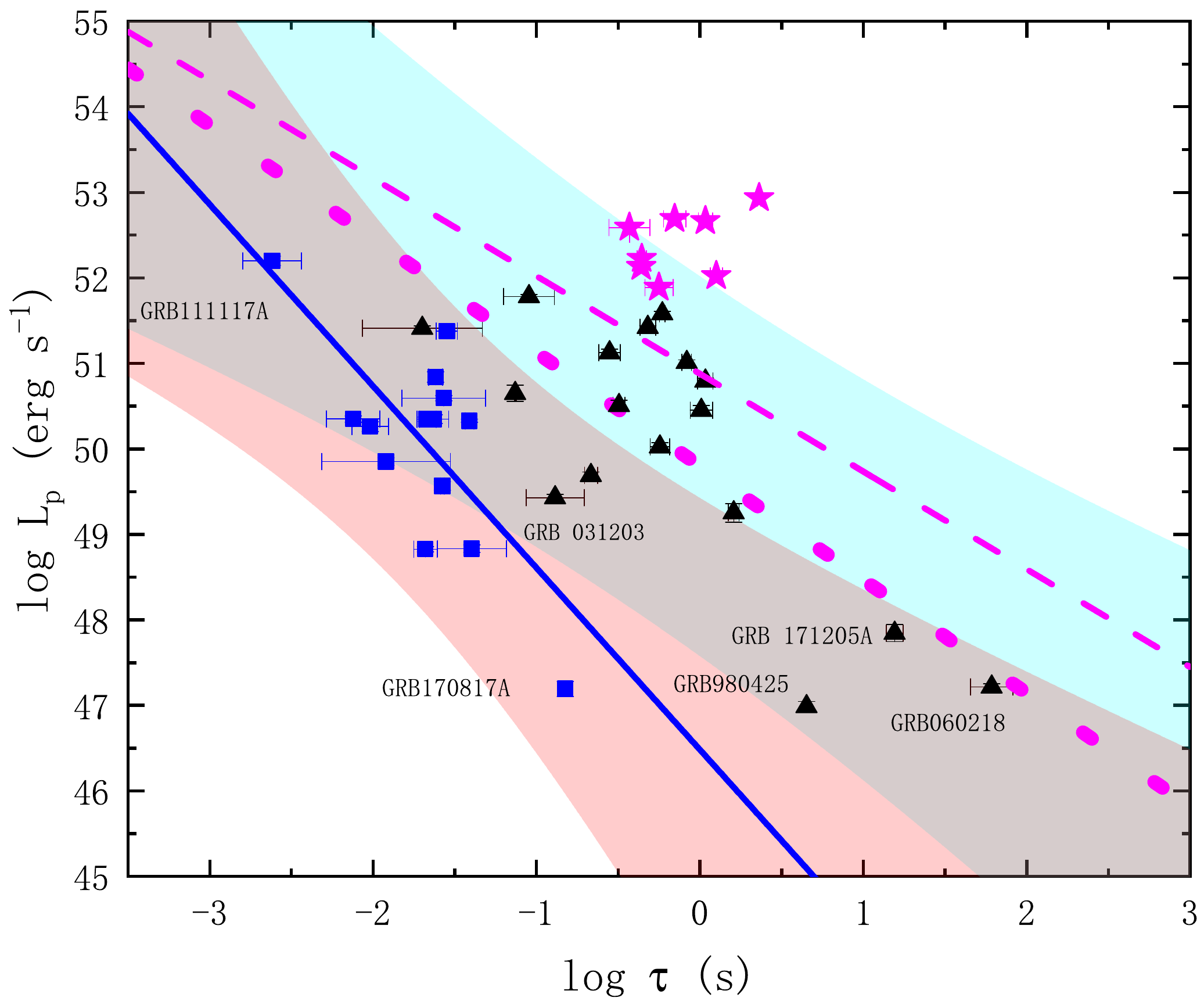}
	\includegraphics[width=\columnwidth]{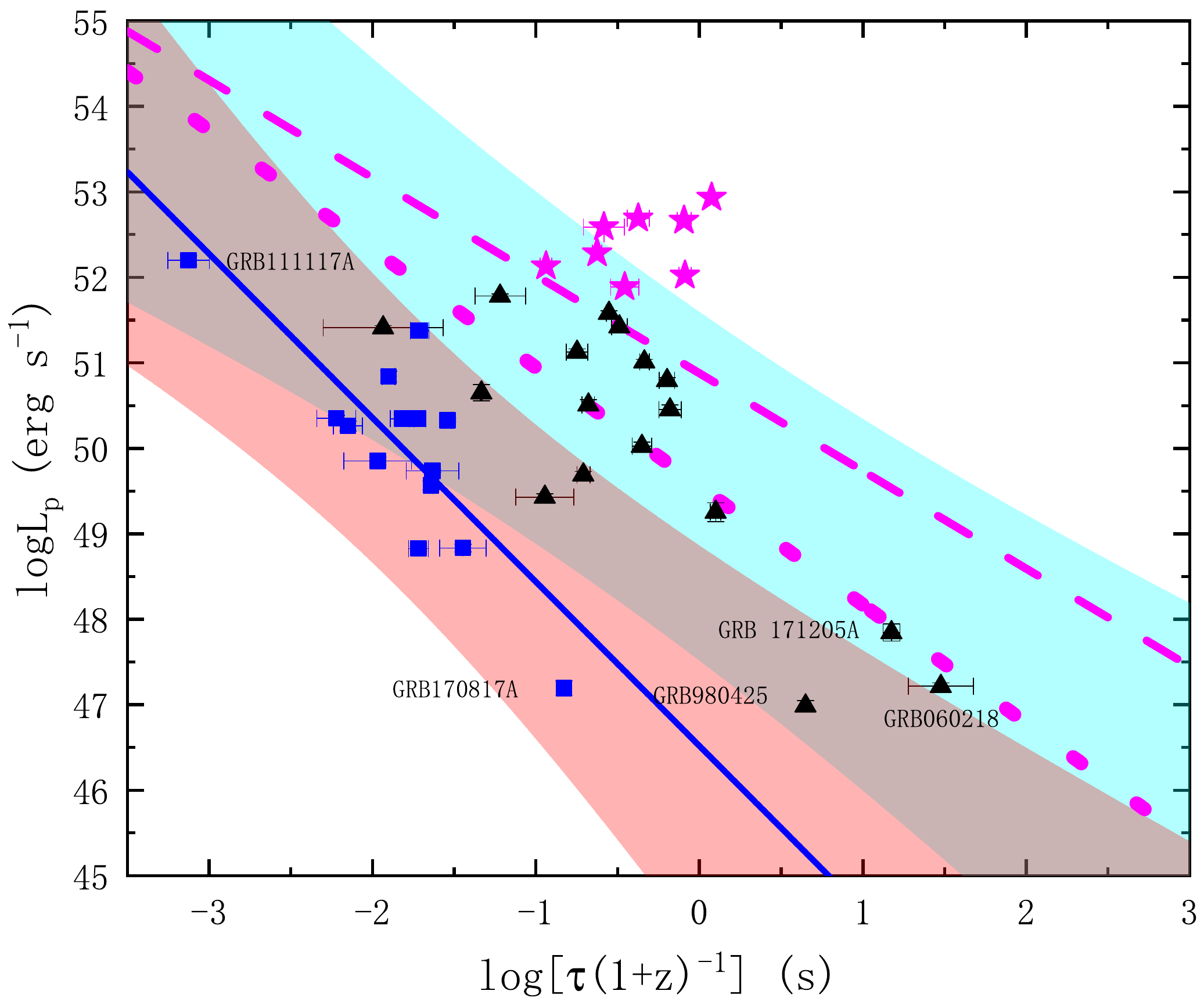}
	\caption{Correlations between the spectral lag and the peak luminosity of 17 SN/GRBs (triangles) and 14 KN/GRBs (squares) in the observer (left panel) and the rest (right panel) frames. The dotted/solid lines stand the best power-law fits to the SN/KN-associated GRB data. The eight SLSN/GRBs marked with magenta stars are not involved in the fitting processes. The dashed lines represent the correlations proposed for normal bursts by \citet{2000ApJ...534..248N}. The light and heavy regions represent the prediction and confidence bands at a level of 2$\sigma$.}	
	\label{fig:lag and lag/1+z}
\end{figure*}
Now let's investigate how the spectral evolution is related to the peak luminosity for the SN- and KN-associated GRBs. As usual, the spectral evolution is described with the time delay or lag of light curves between different energy channels. The \textit{Swift}/BAT GRB light curves of energy channels 1 (15-25 keV) and 3 (50-100 keV) are adopted for the calculation by using the cross-correlation function (CCF) method \citep{1995ApJ...439..307F,1995A&A...300..746C,1997ApJ...486..928B}. Out of 53 SN/GRBs and 15 KN/GRBs, we adopte 16 SN/GRBs and 14 KN/GRBs for the cross-correlation analysis after eight SLSN/GRBs are excluded. Note that the fractions of \textit{Swift}/BAT bursts in both selected samples are more than 90\%, which guarantees our calculations to avoid the instrumental bias and sample selection effect in a certain sense. Subsequently, we use a Gaussian function to fit the CCF curve and get the spectral lag. Figure \ref{fig:lag and lag/1+z} shows the correlations of the spectral lag with the peak luminosity for two kinds of GRBs. The best fits with a power-law function in the left panel give
\begin{equation}
\texttt{log}L_p=(-1.43\pm0.33)\texttt{log}\tau+(49.57\pm0.27),
\end{equation}
with a Pearson's coefficient of $r=-0.76$ for the SN/GRBs and
\begin{equation}
\texttt{log}L_p=(-2.17\pm0.57)\texttt{log}\tau+(46.34\pm0.99),
\end{equation}
with a Pearson's coefficient of $r=-0.74$ for the KN/GRBs in the observer frame. Meanwhile, we also exhibit the correlations between the spectral lag and the peak luminosity in the right panel and obtain the following best fits of
\begin{equation}
\texttt{log}L_p=(-1.42\pm0.28)\texttt{log}[\tau/(1+z)]+(49.39\pm0.26),
\end{equation}
with a Pearson's coefficient of $r=-0.80$ for the SN/GRBs and
\begin{equation}
\texttt{log}L_p=(-1.92\pm0.40)\texttt{log}[\tau/(1+z)]+(46.57\pm0.77),
\end{equation}
with a Pearson's coefficient of $r=-0.80$ for the KN/GRBs in the rest frame. Importantly, this demonstrates that the peak luminosity is anti-correlated with the time lag not only for normal GRBs as suggested by \cite{2000ApJ...534..248N} but also for the SN- and KN-associated GRBs. Moreover, their power-law indices are significantly distinct and all the three relations are somewhat overlapped, indicating that the diverse types of GRBs could share the same observational features. We exclued GRBs 041006, 100418A, 180728A and 190829A because of their negative lags. Interestingly, the power-law index of KN/GRBs is consistent with $-1.79\pm0.03$ of the luminosity-lag relation for the complete Swift LGRBs with positive lags only in the rest frame \citep{2015MNRAS.446.1129B}.

We caution that eight SLSN/GRBs with luminosity larger than $L_{b2}\simeq6.5\times{10^{51}}$erg s$^{-1}$ are not involved in our fitting processes. It is interestingly found that the peak luminosities of SLSN/GRBs are independent of the spectral lags, which may imply that they should represent a peculiar subclass of SN/GRBs. It is worth pointing out that GRB 980425 as an outlier of the Norris's $L_p-\tau$ relation complies with our new-finding relationships of SN/GRBs in the observer and the source frames as well. Some former studies argued that the anti-correlation between $L_p$ and $\tau$ may be directly related to the initial bulk Lorentz factors ${\Gamma _0}$ \citep{2000ApJ...544L.115S,2001ApJ...563L.123S,2003MNRAS.342..587D,2006MNRAS.373..729Z}. This reflects the complexity and diversity of dynamics, components and geometry of the relativistic outflows launched from different kinds of GRBs.

Noticeably, two kinds of GRBs are somewhat overlapped in Figure \ref{fig:lag and lag/1+z} although both $L_p$ and $\tau$ of KN/GRBs are on average smaller than the corresponding values of SN/GRBs and they are all located at their own 2$\sigma$ prediction ranges. Interestingly, 7 SN/GRBs and 9 KN/GRBs conform to both the luminosity-lag relations in the observer frames simultaneously, with corresponding fractions of $\sim$41.18\% and $\sim$64.29\%, correspondingly. There are 5 SN/GRBs and 6 KN/GRBs following both the two relations in the rest frame, of which the corresponding fractions are $\sim$29.41\% and $\sim$42.86\%. More interestingly, GRB 171205A as a long burst associated with Type Ic SN 2017iuk \citep{2017GCN.22180....1I,2018ApJ...867..147W,2018A&A...619A..66D} perfectly conforms to the anti-correlation of luminosity with time lag of SN/GRBs and  lies on the edge of 2$\sigma$ regions of the $L_p-\tau$ relation for the KN/GRBs. This demonstrates that the low-luminosity GRB 171205A may have the similar origin as other KN/GRBs. In practice, \cite{2022NewA...9701889K} deduced from the analysis of X-ray plateau that the central engine of GRB 171205A should be a magnetar. In addition, the flux property of its host galaxy is similar to the other two SN/GRBs 980425 and 031203 (Li et al. 2022).

\subsubsection{The Amati relation}
\begin{figure}
	\centering
	\includegraphics[width=1\columnwidth]{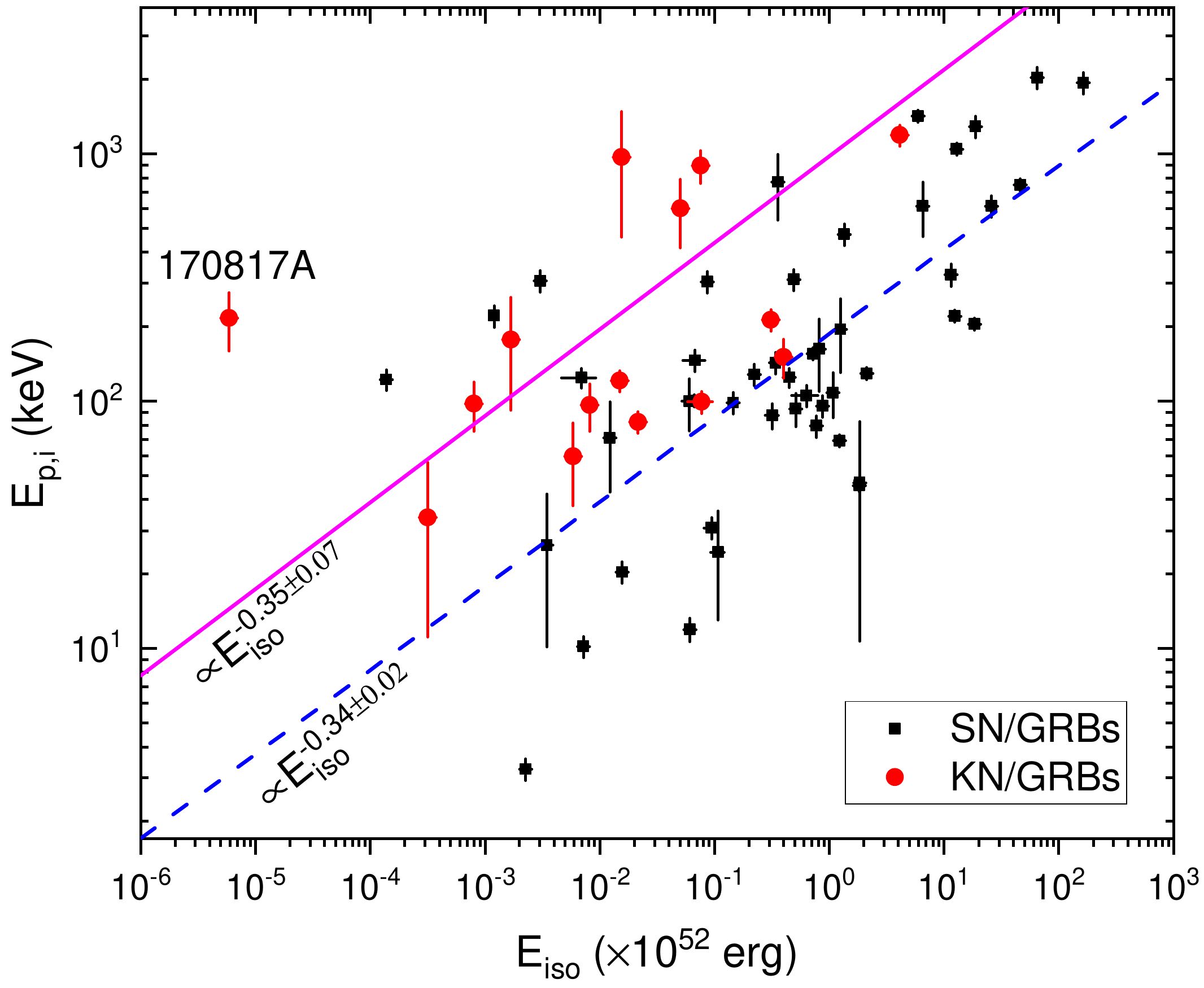}
   	\caption{Relation between $E_{p,i}$ and $E_{iso}$ for SN/GRBs (filled squares) and KN/GRBs (filled circles). The best fits to the Amati relations of short (solid line) and long (dashed line) bursts are taken from
\citet{2018PASP..130e4202Z}.}
	\label{fig:Amati}
\end{figure}

As the number of SN/GRBs and KN/GRBs increases, it becomes more and more feasible to check whether they have the similar Amati energy relations of $E_{p,i}\propto E_{iso}^\eta$ \citep{2002A&A...390...81A} between the intrinsic peak energy $E_{p,i}=E_p(1+z)$ and the isotropic energy output $E_{iso}=4\pi D^2_{L}S_{bolo}(1+z)^{-1}$, in which $D_{L}$ and $S_{bolo}$ respectively denote the luminosity distance and the bolometric luminosity of a burst \citep[see details in][]{2018PASP..130e4202Z}. We perform a linear correlation analysis and obtain the Pearson coefficients to be $r=0.75$ with a P-value of $1.9\times10^{-9}$ for SN/GRBs and $r=0.61$ with a P-value of $1.5\times10^{-2}$ for KN/GRBs, indicating $E_{p,i}$ and $E_{iso}$ of both types of GRBs are positively correlated. We further compare the observations of two kind bursts with the derived Amati relations by \cite{2018PASP..130e4202Z} in Figure \ref{fig:Amati}, where it can be clearly found that the SN- and KN-associated GRBs roughly coincide with the Amati relations of long and short GRBs, respectively. However, it is worth noting that GRB 170817A as a KN-associator of AT 2017gfo \citep{2017Sci...358.1556C,2017ApJ...848L..12A} is an outlier of the Amati relation of not only KN/GRBs but also SN/GRBs. This may be attributed to the off-axis effect on the resulting energy relations with power law indices smaller than the theoretical value of $\eta=0.5$ given by the synchrotron radiation mechanism \cite[e.g.][]{2022arXiv221104727X} and even the observed index of $\eta\simeq0.35$ for short GRBs \citep{2018PASP..130e4202Z}. Furthermore, we notice that most SN/GRBs and KN/GRBs are heavily overlapped in the plot of $E_{p,i}$ against $E_{iso}$. The larger scatter in the plot may reflect the diversity and complexity of both classes of GRBs. \textbf{In addition, more parameters of prompt $\gamma$-rays of SN/KN GRBs are compared with a K-S test and listed in Table \ref{tab:7}.}
\subsection{Properties of SN/KN-connected GRB X-ray afterglows}
\subsubsection{X-ray afterglow morphology}
\label{subsec: all-sky}
In this section, we investigate the differences in X-ray light curves between SN/GRBs and KN/GRBs. For this purpose, we select the two kinds of GRB samples of \textit{Swift}/XRT X-ray afterglows with enough data points from ${10}$ to ${10^6}$ seconds since trigger to discriminate the temporal profiles in X-rays. Following the selection criteria defined by \cite{2008ApJ...679..570S}, we categorize these X-ray afterglows into three cases: (1) Those exhibiting a flare-like bump; (2) those with a plateau segment; and (3) those with a steep decline after prompt $\gamma$-ray emissions. The isotropic X-ray luminosity ${L_{\rm X}}$ in the energy band of 0.3-10 keV at the observation time $t$ is determined by the formula \citep{2019ApJS..245....1T,2021AAS...23723306D}
\begin{equation}
{L_{\rm X,iso}(t)} = 4\pi D_{\rm L}^2{F_{\rm X}(t)}(1+z)^{-(2-\Gamma_X)}\ \ \rm erg\ s^{-1}
\end{equation}
where $F_{\rm X}(t)$ is the observed flux at a given time $t$ and $\Gamma_X$ denotes the photon spectral index obtained from
the official \textit{Swift} GRB table\footnote{https://swift.gsfc.nasa.gov/archive/grb\_table.html/}. The K-correction factor of X-rays is determined by $K_{c,X}=(1+z)^{\Gamma_X-1}$.

\begin{figure*}
	\centering
	\includegraphics[width=\columnwidth]{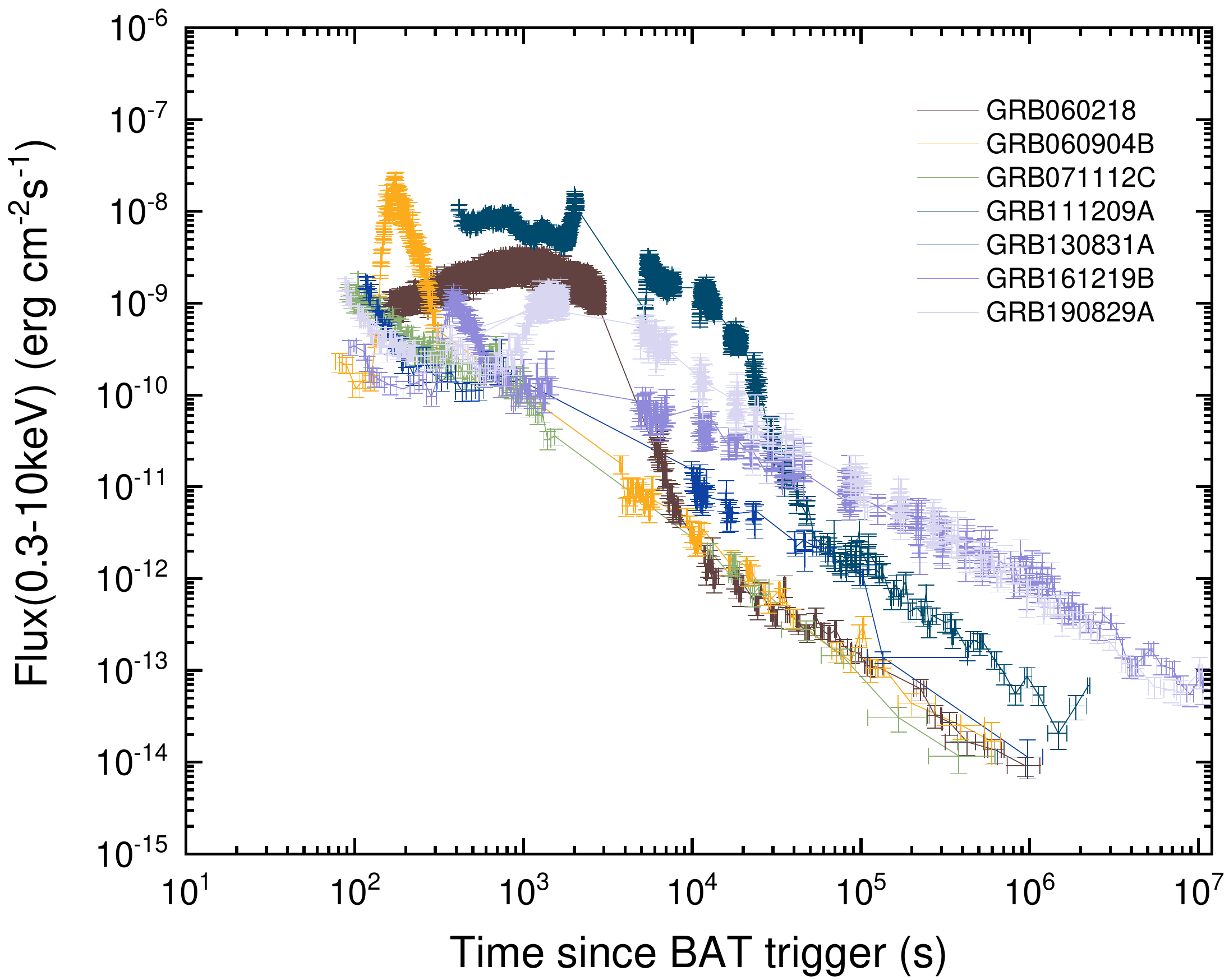}
	\includegraphics[width=\columnwidth]{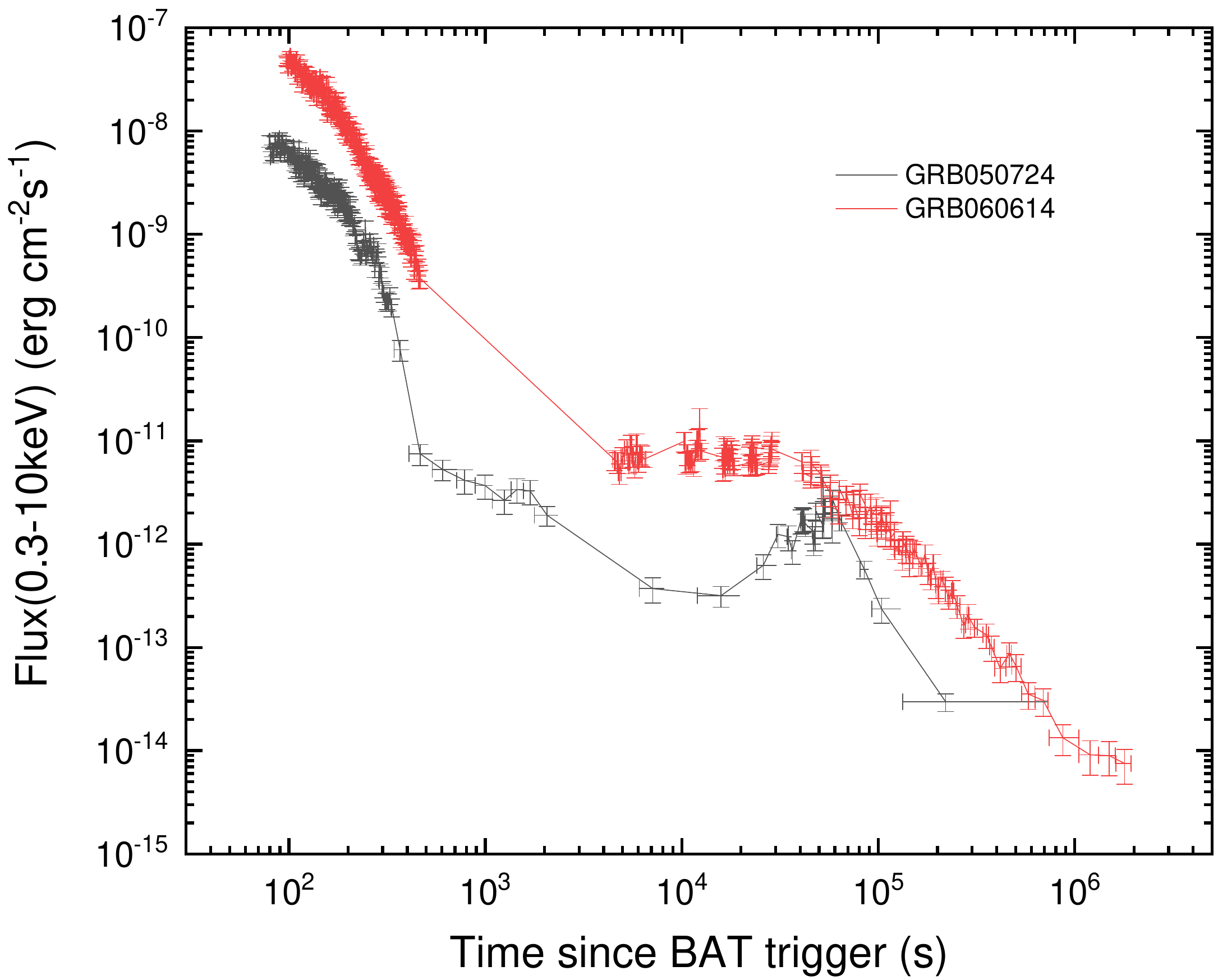}
	\includegraphics[width=\columnwidth]{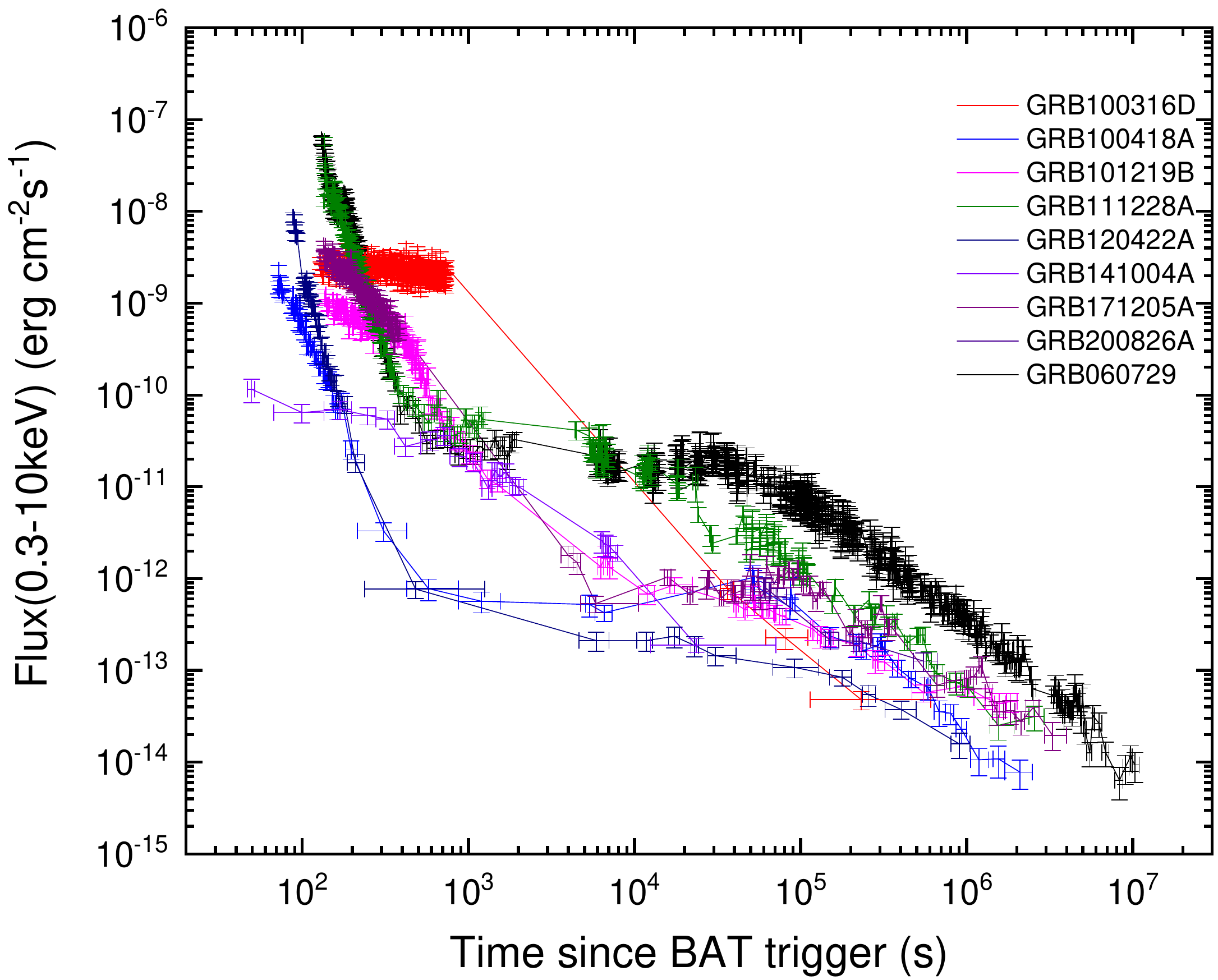}
	\includegraphics[width=\columnwidth]{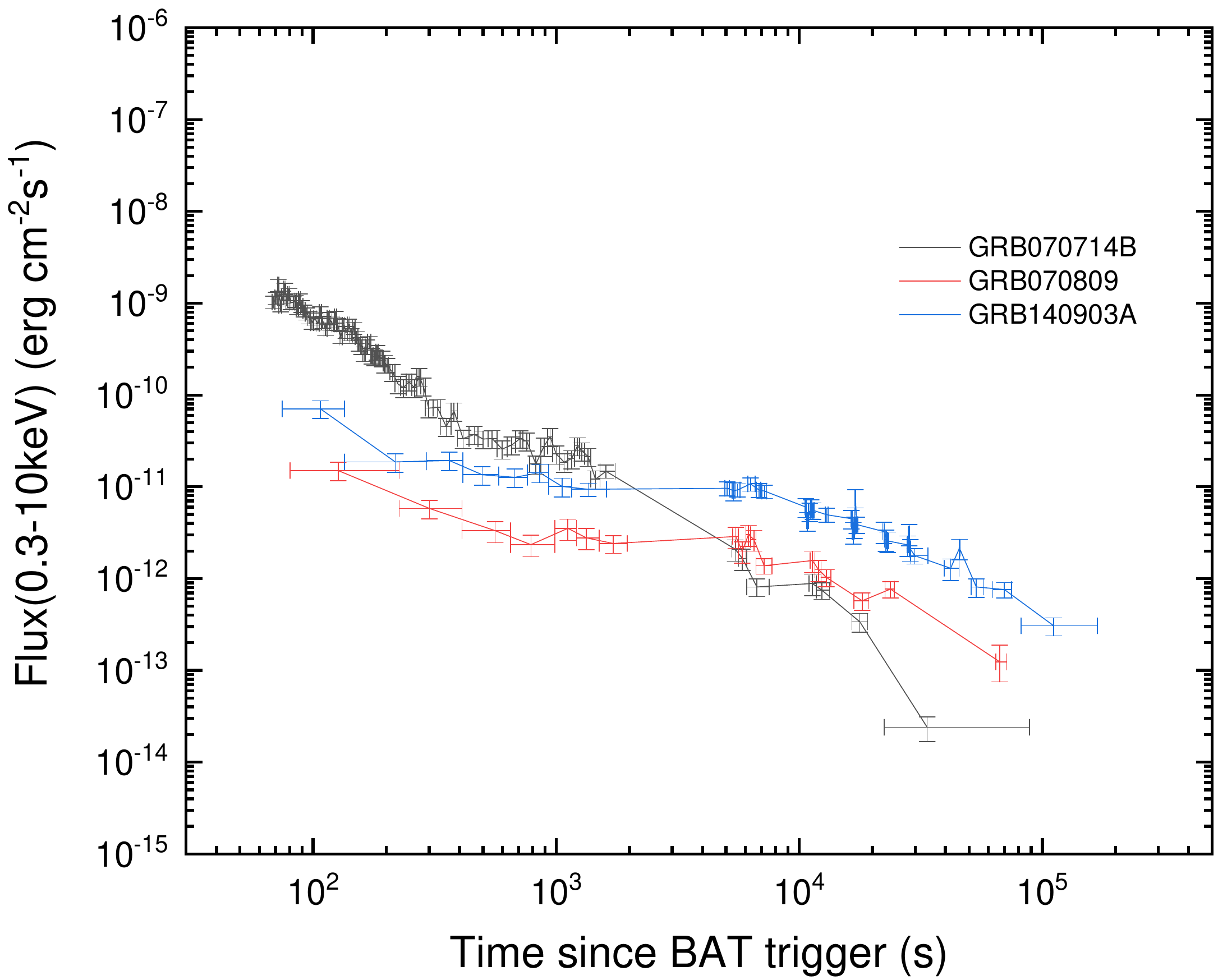}
	\includegraphics[width=\columnwidth]{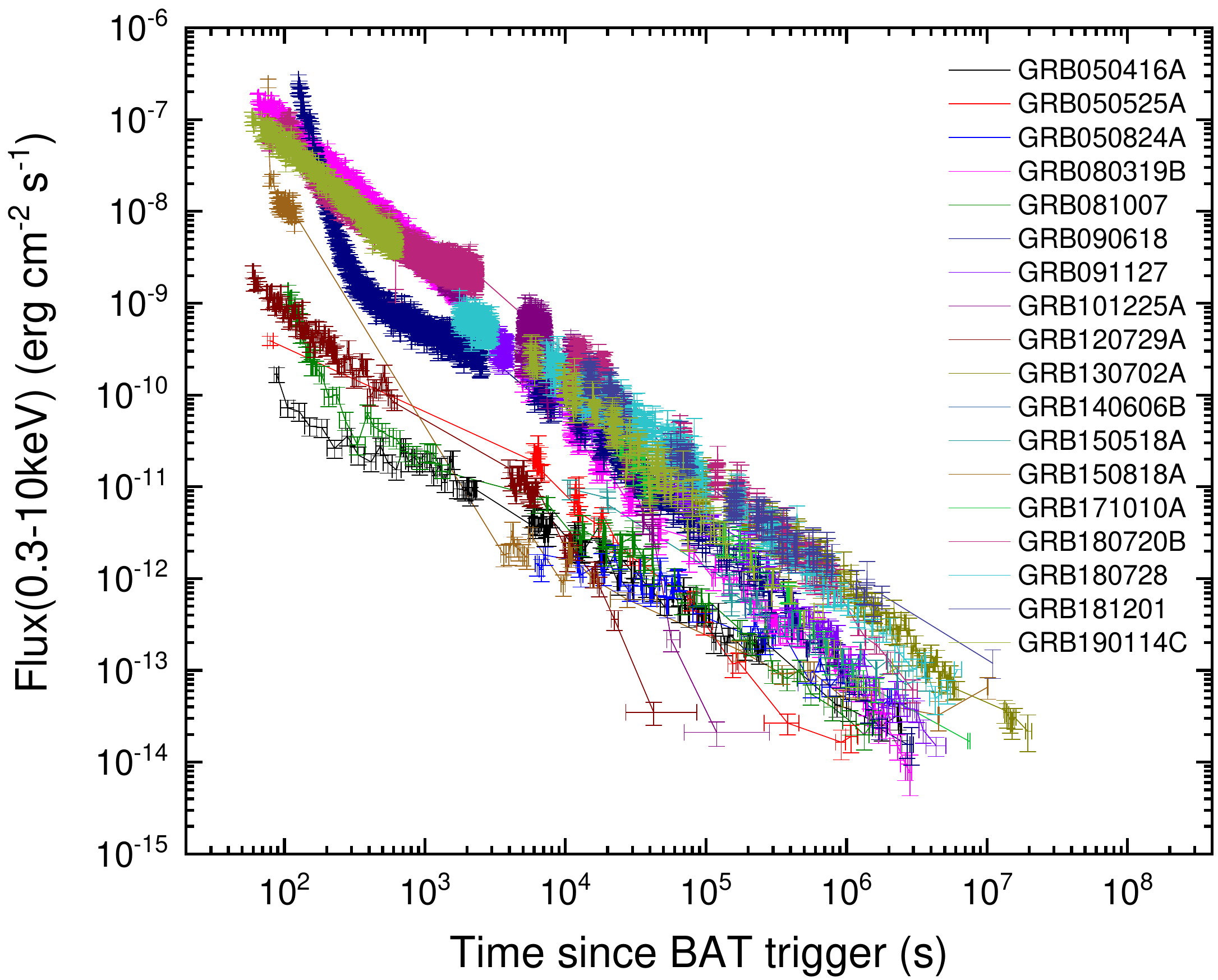}
	\includegraphics[width=\columnwidth]{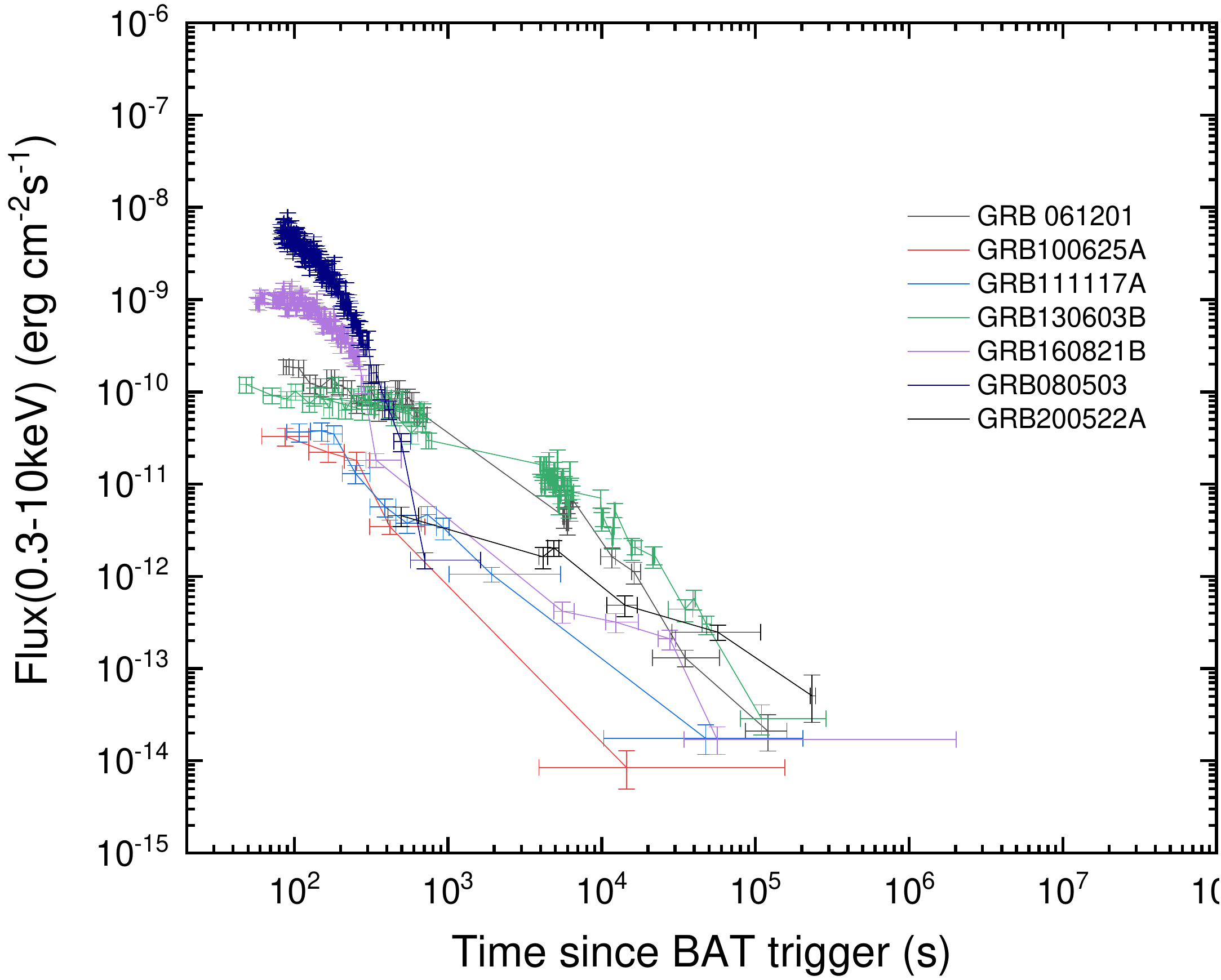}
	\caption{\textit{Swift}/XRT (0.3-10 keV) X-ray afterglows of case 1, case 2, and case 3 in our samples of SN/GRBs (left panels) and KN/GRBs (right panels). Different bursts are symbolized with distinct colors.}
	\label{fig:fluxcomposite}
\end{figure*}

\begin{figure*}
	\centering
	\includegraphics[width=\columnwidth]{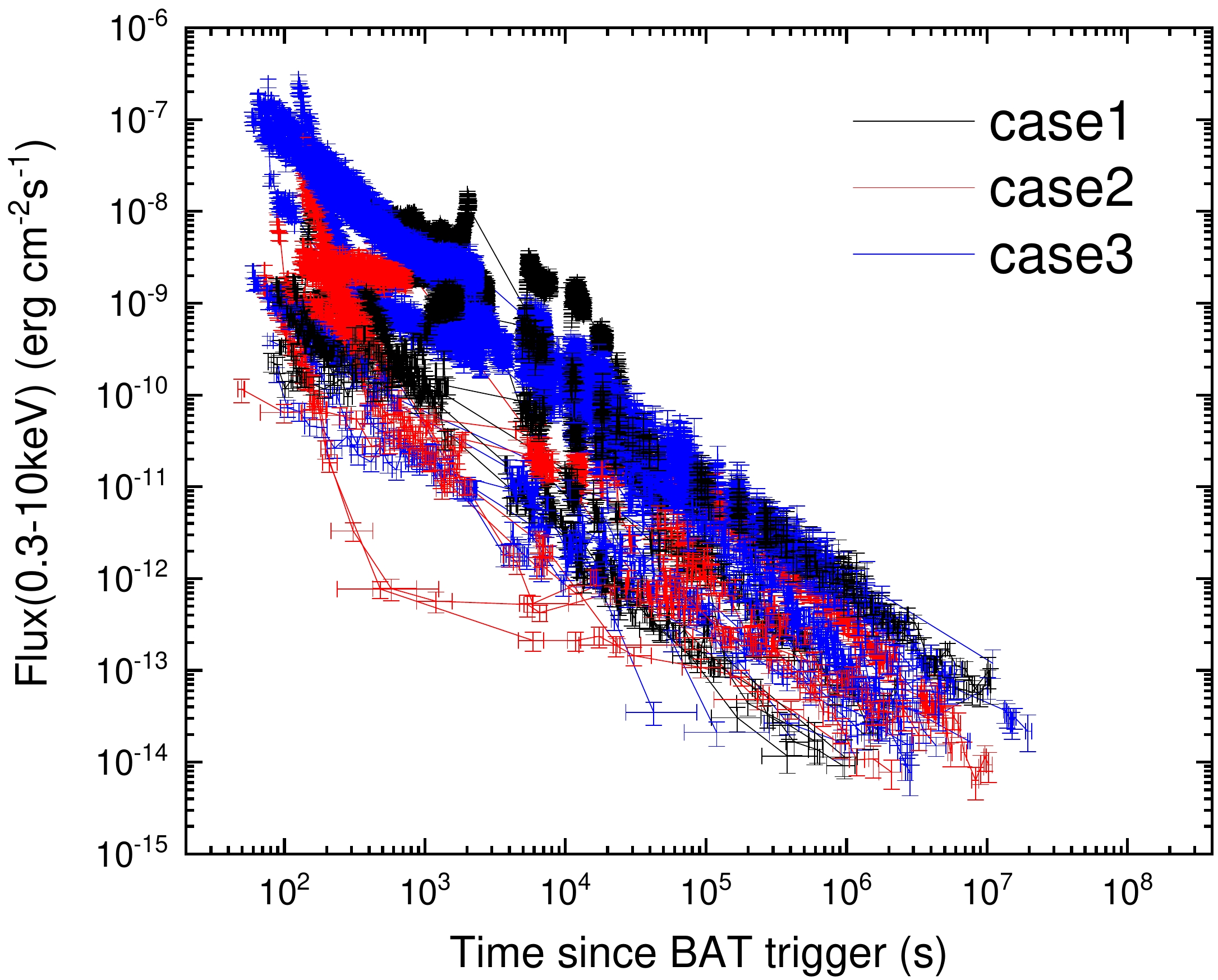}
	\includegraphics[width=\columnwidth]{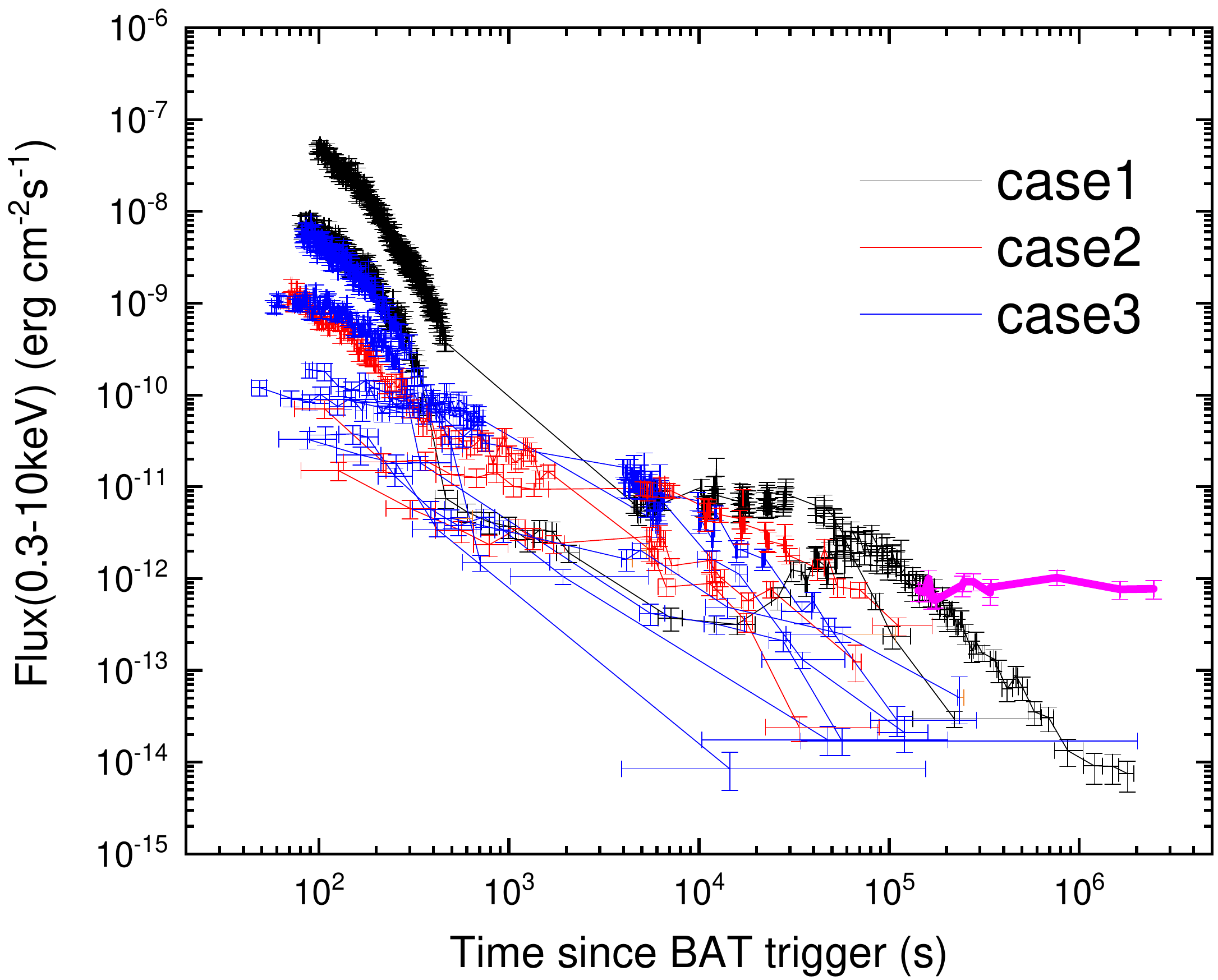}
	\caption{\textit{Swift}/XRT (0.3-10 keV) X-ray afterglows of the total SN/GRB (left panel) and KN/GRB (right panel) samples. GRB 150101B is excluded owing to its very odd light curve. Different cases are distinguished by the symbol colors.}
	\label{fig:fluxtotal}
\end{figure*}

\begin{figure*}
	\centering
	\includegraphics[width=\columnwidth]{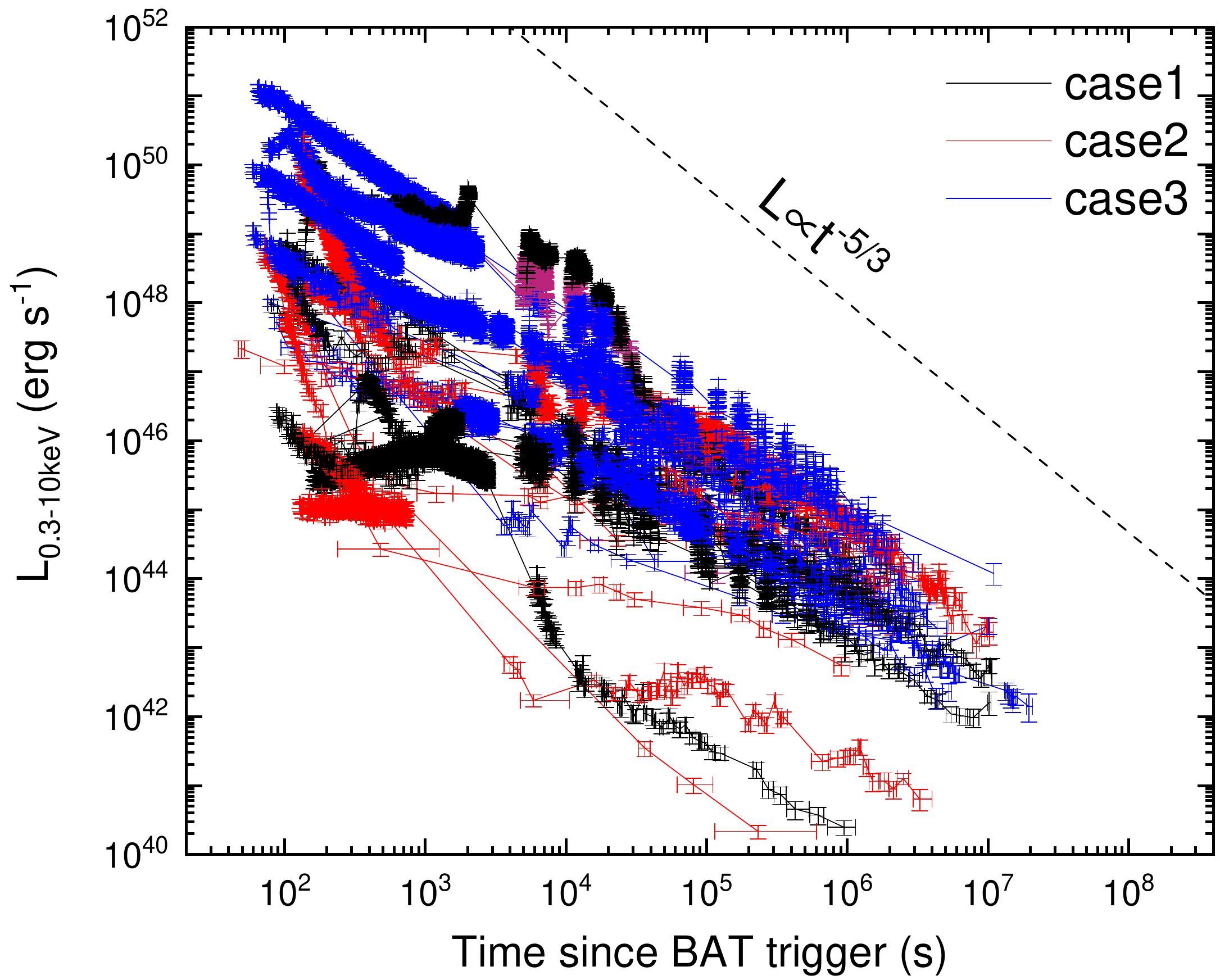}
	\includegraphics[width=\columnwidth]{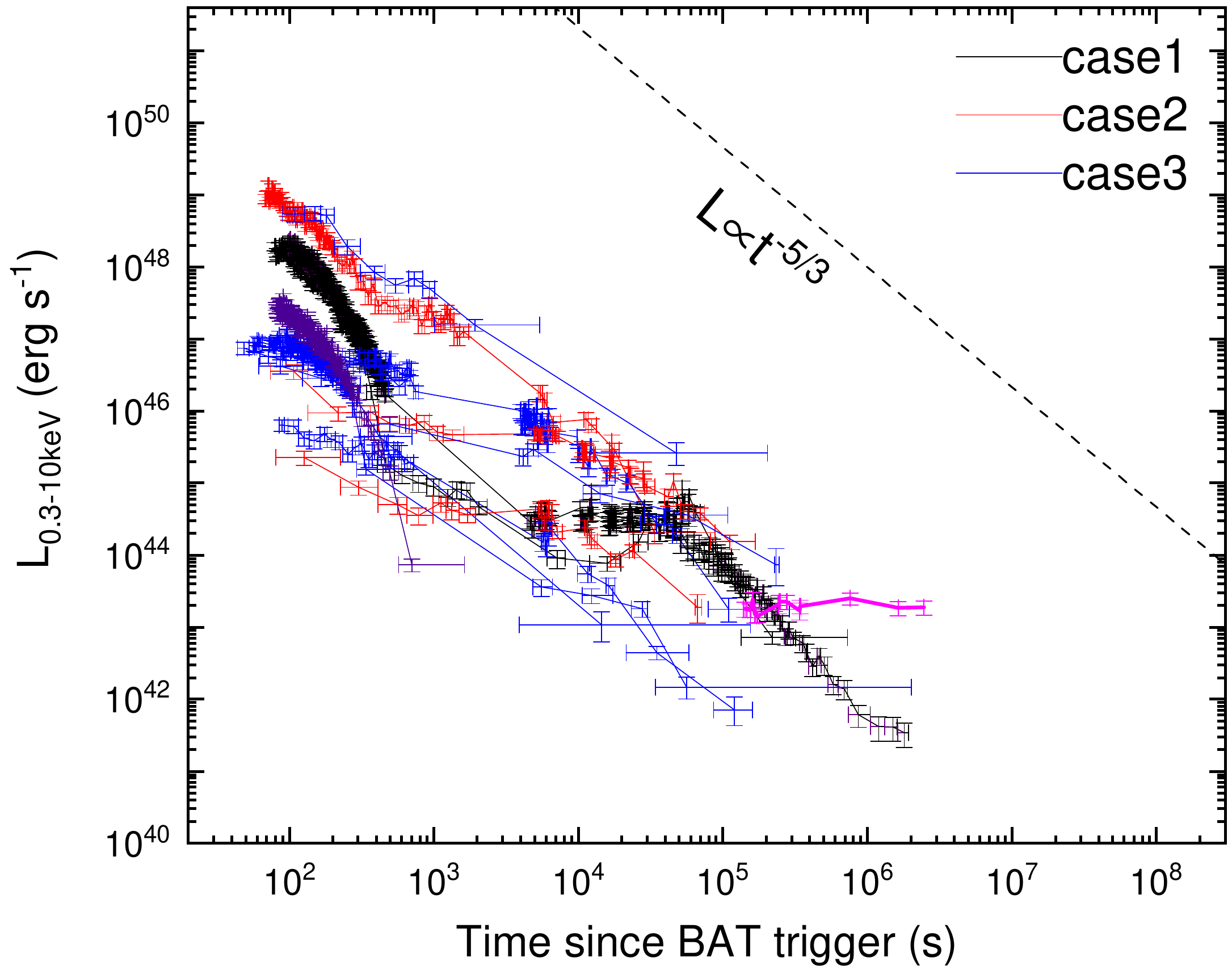}
	\caption{The luminosities are plotted against the observation times for the \textit{Swift}/XRT X-ray afterglows of SN/GRBs (left panel) and KN/GRBs (right panel) within the energy range of 0.3-10 keV. GRB 150101B is also excluded. The dashed lines show the power-law relation of the total luminosity of the Blandford-Znajek (BZ) jet from a Black Hole (BH) with time with an index of -5/3. All the symbols are the same as Figure \ref{fig:fluxtotal}.}
	\label{fig:xraylum}
\end{figure*}

\begin{figure*}
	\centering
    \includegraphics[width=\columnwidth]{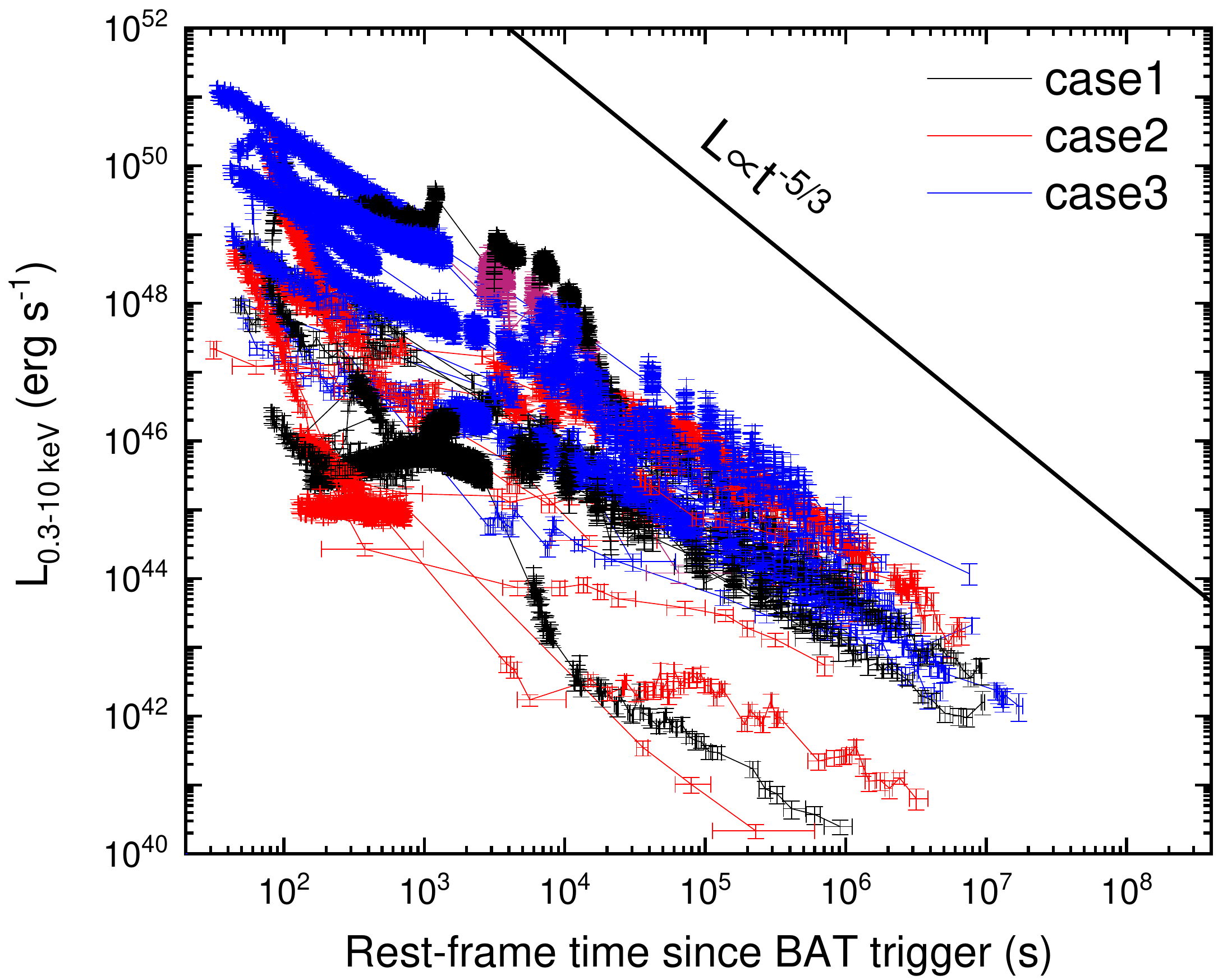}
	\includegraphics[width=\columnwidth]{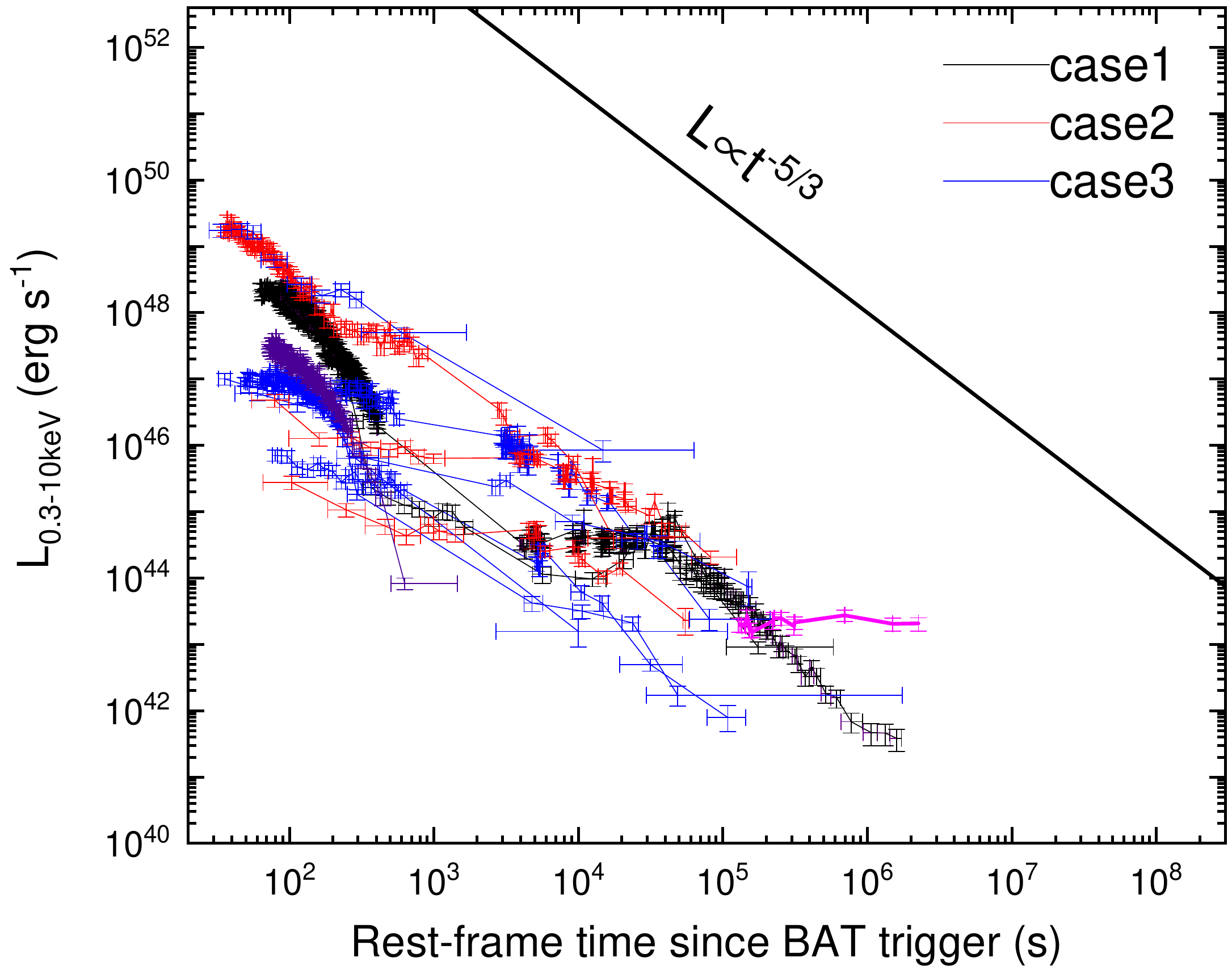}
	\caption{The luminosities are plotted against the intrinsic times for the \textit{Swift}/XRT X-ray afterglows of SN/GRBs (left panel) and KN/GRBs (right panel) within the energy range of 0.3-10 keV. GRB 150101B is also excluded. The solid lines show the power-law relation of the total luminosity of the Blandford-Znajek (BZ) jet from a Black Hole (BH) with time with an index of -5/3. All the symbols are the same as Figure \ref{fig:fluxtotal}.}
	\label{fig:xraylumin}
\end{figure*}

Figure \ref{fig:fluxcomposite} shows the X-ray light curves of SN/GRBs and KN/GRBs in each case as mentioned above. The total temporal profiles of X-ray afterglows in 0.3-10 keV are compared in Figure \ref{fig:fluxtotal}, from which we can find that the X-ray rebrightening of SN/GRBs occurs earlier than that of KN/GRBs. In particular, the bump components of SN/GRBs in case 1 generally appear within the time interval of $10^{2}-10^{4}$ s after bursts, while the rebrightening emissions of KN/GRBs in case 1 arise after $10^4$ s since triggers. For case 2, the X-ray profiles of KN/GRBs are evidently flatter than those of SN/GRBs and their plateaus end relatively earlier. The temporal indexes of X-ray light curves of SN/GRBs and KN/GRBs in case 3 are very similar, but the averaged brightness of SN/GRBs is roughly one order of magnitude higher than that of KN/GRBs at the same stage. In addition, we depict the relationships of luminosity with the observation time in Figure \ref{fig:xraylum} and the intrinsic time in Figure \ref{fig:xraylumin} to show the variable behaviors of X-ray light curves of these SN- and KN-selected bursts. We can see that the averaged luminosity of SN/GRBs is about two orders of magnitude larger than that of KN/GRBs, which is quite similar to the discrepancy of luminosities between LGRBs \citep{2007A&A...469..379E,2009MNRAS.397.1177E} and SGRBs \citep{2011ApJ...734...96K,2014ARA&A..52...43B,2018A&A...616A..48S}. It is noticeable that the luminosities of SN/GRBs compared with the KN/GRBs are distributed in a larger scope. In addition, we compare the descending trend of the SN/KN luminosities with the total luminosity of the Blandford-Znajek (BZ) jet from a Black Hole (BH) and find that they exhibit a similar behavior as a whole. This implies that the BZ mechanism could be ubiquitous among different kinds of SN/KN GRBs.

\subsubsection{Relation of X-ray luminosity versus plateau timescale}
\begin{figure*}
	\centering
	\includegraphics[width=\columnwidth]{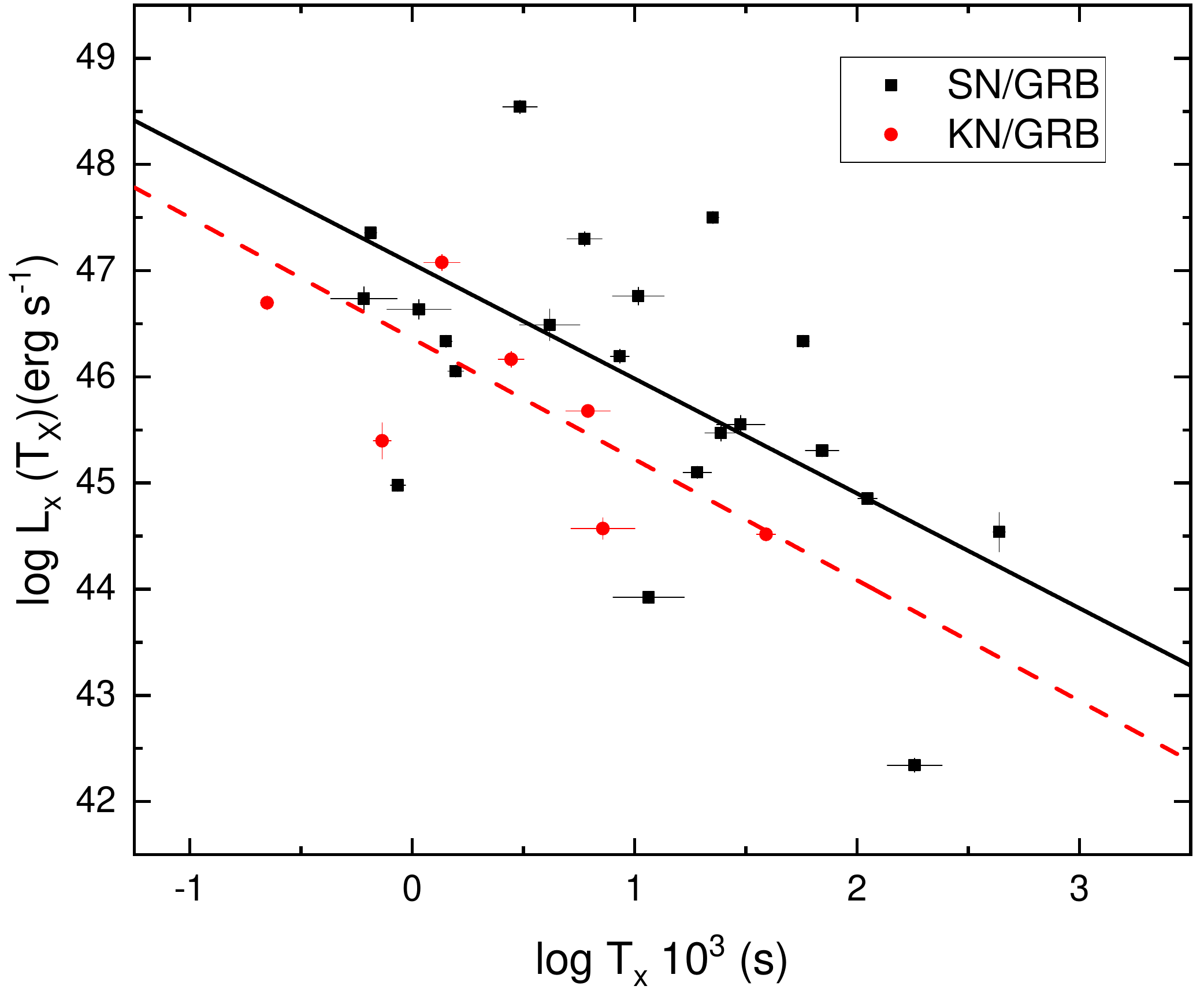}
	\includegraphics[width=\columnwidth]{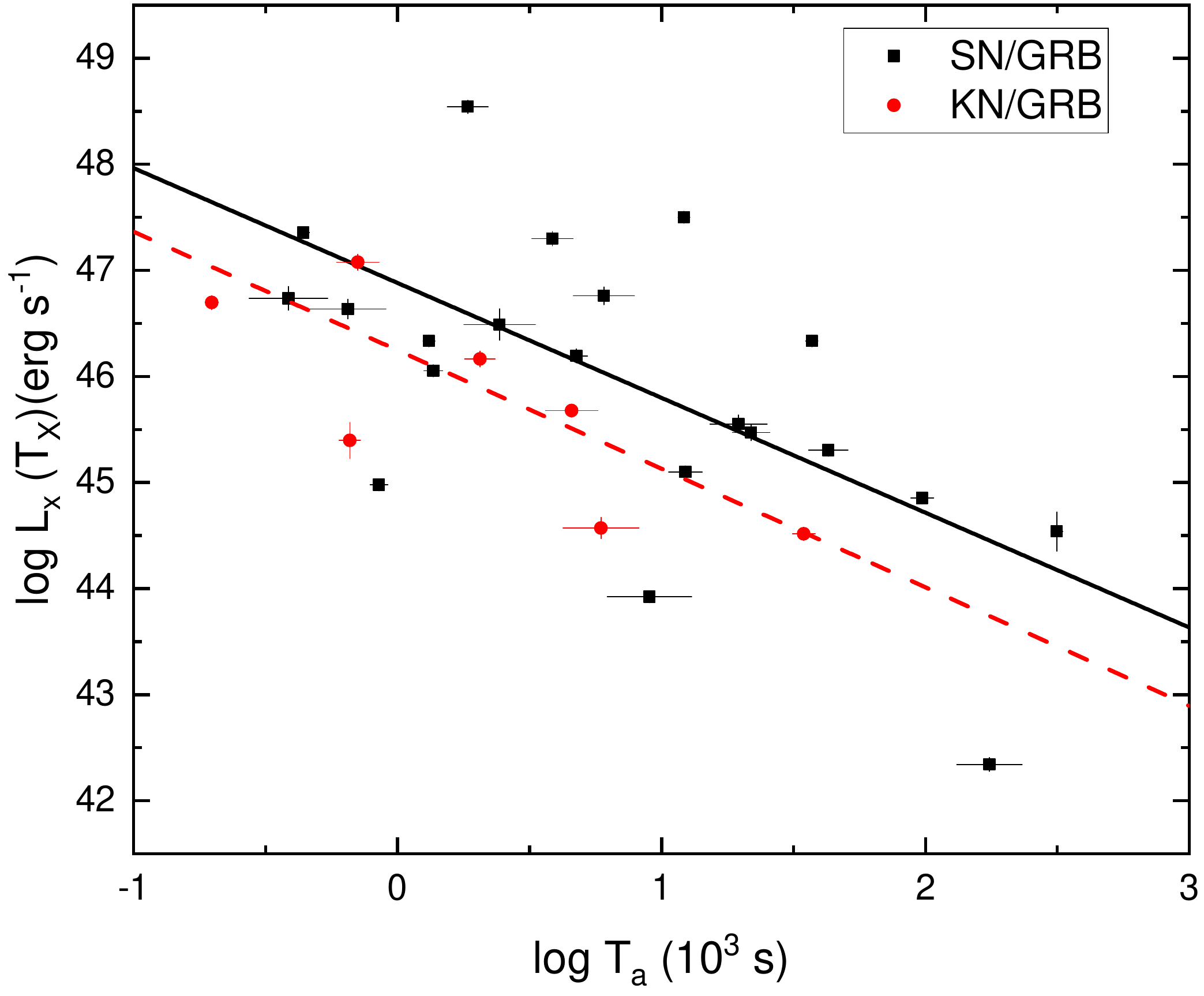}
	\caption{Dependence of the X-ray luminosity on the end time of X-ray plateaus for both 21 SN/GRBs and 7 KN/GRBs in the observer (left panel) and rest (right panel) frames.}
	\label{fig:Lx-Tx}
\end{figure*}

To examine the dependence of X-ray luminosity on the end time ($T_X$) of plateau, we select 21 SN/GRBs and 7 KN/GRBs with well-fitted X-ray plateaus by the smoothly broken power-law function \cite[e.g.][]{2009MNRAS.397.1177E,2019ApJS..245....1T} as follows
\begin{equation}
F_X(t)=F_{X0}\big[\big(\frac{t}{T_X}\big)^{\alpha_1\omega}+\big(\frac{t}{T_X}\big)^{\alpha_2\omega}\big]^{-1/\omega},
\end{equation}
where the smoothness parameter $\omega=3$ has been assumed in our calculation \citep{2019ApJS..245....1T}. Thus the X-ray luminosity at the time of $T_X$ is given by
\begin{equation}
{L_{\rm X}(T_X)}=4\pi D_{\rm L}^2F_{X0}2^{-1/\omega}(1+z)^{(\Gamma_{X}-2)}  \ \rm erg\ s^{-1}.
\end{equation}
Table \ref{tab:8} displays the best fitted parameters of X-ray afterglow plateaus. The luminosity-time relation is depicted in Figure \ref{fig:Lx-Tx}, from which we see that $L_X$ and $T_X$ can be fitted by assuming a power-law form $L_X\propto T_X^\xi$, where the power law indices are $\xi=-1.08\pm0.24$ for SN/GRBs and $\xi=-1.14\pm0.21$ for KN/GRBs. Considering the intrinsic plateau time as $T_a=T_X/(1+z)$, one can suppose $L_X\propto T_a^\zeta$ and get the best fits $L_{X,SN}\propto T_a^{-1.08\pm0.22}$ for SN/GRBs and $L_{X,KN}\propto T_a^{-1.12\pm0.17}$ for KN/GRBs in the $L_X-T_a$ plane. It is excitingly found that the power-law indices of the X-ray luminosity versus the plateau time of SN/GRBs and KN/GRBs are consistent with each other regardless of their significant difference in peak luminosity of prompt $\gamma$-rays as displayed in Figure \ref{fig:lpdis}. Surprisingly, the power indices inferred in the work are identical to the values of $\zeta\simeq-1.10\pm0.20$ \citep{2019ApJS..245....1T} and $\zeta\simeq-1.06_{-0.28}^{+0.27}$ \citep{2010ApJ...722L.215D} for long GRBs. \cite{2017A&A...600A..98D} obtained $\zeta=-1.5\pm0.3$ with 19 SN/GRB candidates that is slightly steeper than that of normal long bursts. They speculated that the SN/GRBs might not require a standard energy injection in the plateau phase. It is also possibly due to the statistical fluctuation since only 7 out 19 bursts are confirmed to associate with SNe in their sample. In theory, the luminosity and plateau duration of X-ray afterglows can be naturally correlated by $L_X\sim T_a^{-1}$ or $L_X\sim T_X^{-1}$ if only the process of energy injection into forward shock from the compact objects and the interaction between the shock and the circumstellar medium are taken into account \citep{2014MNRAS.443.1779R}. Our findings suggest that the power-law index $\zeta\simeq-1$ of both SN/GRBs and KN/GRBs with X-ray plateaus can be ubiquitously interpreted in the same radiation mechanism in despite of distinct progenitors or central engines.

\subsubsection{The $E_p$ distributions of GRBs with different X-ray patterns}
\begin{figure}
	\centering
\includegraphics[width=\columnwidth]{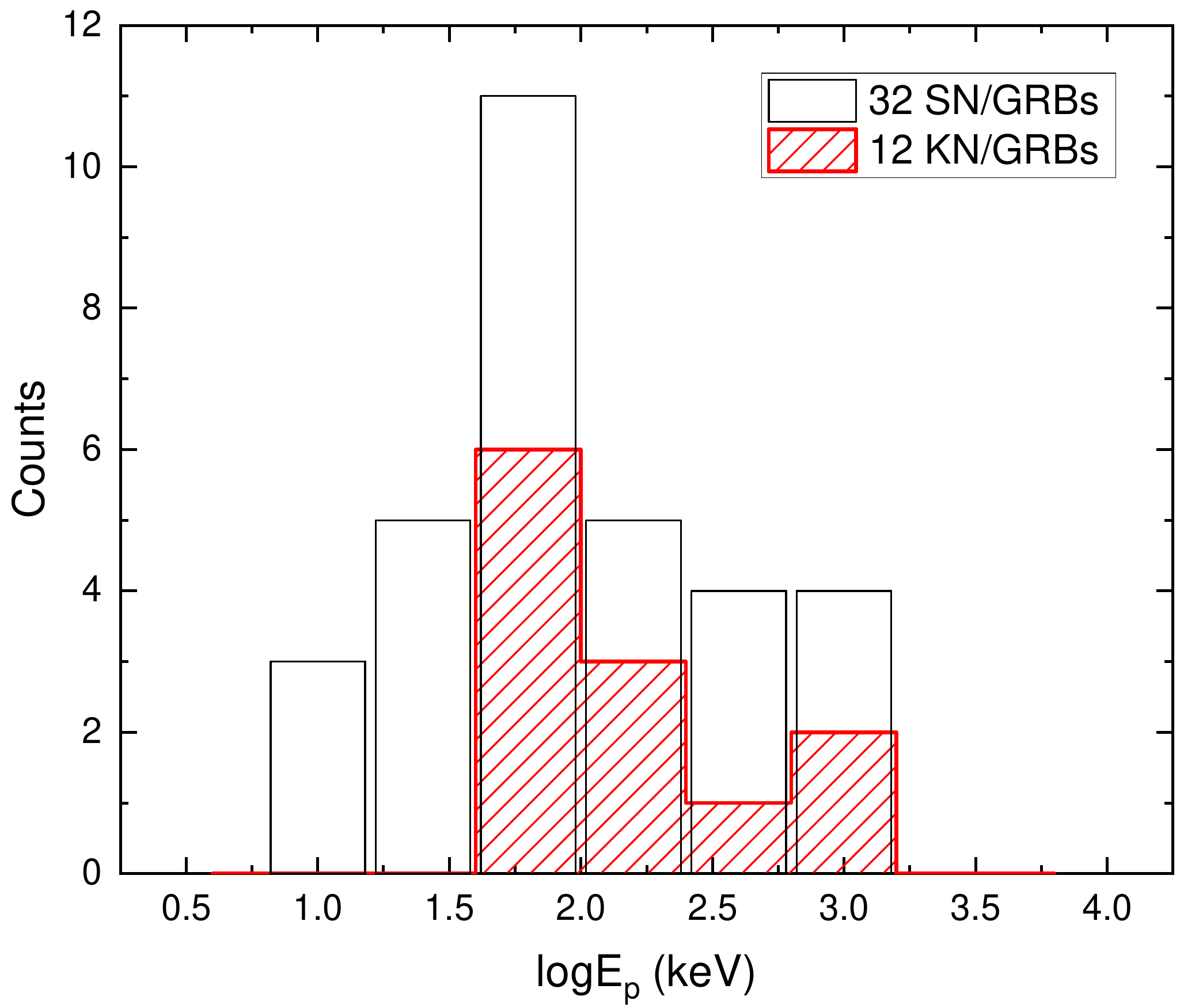}
\caption{The histograms of peak energy $E_{\rm p}$ of 32 SN/GRBs and 12 KN/GRBs with X-ray detections. Note that different subclasses are not identified owing to the limited numbers of KN/GRBs in cases 1 and 2.}
	\label{fig:logep}
\end{figure}

The peak energy $E_p$ as a crucial parameter of the $\nu F_{\nu}$ spectra can provide the spectral properties of GRBs. Previous investigations manifest that the typical $E_p$ is distributed over a broad energy range from a few keV to MeV \cite[e.g.][]{2000ApJS..126...19P,Sakamoto2004}. Usually, brighter GRBs have larger energy outputs, peaking around hundreds of keV to sub-MeV \citep{Preece2016}. This is similar to the $E_p$ distribution of short faint GRBs detected by Konus-Wind \citep{Svinkin2016}. Furthermore, the $E_p$ distributions are influenced by the energy bands of the given detectors. For example, \cite{Zhang2020} obtained that the short and long \textit{Swift} GRBs averagely have the comparable peak energy of $\sim90$ keV, which is approximately half of the typical $E_p\simeq200$ keV reported for the CGRO/BATSE or \textit{Fermi}/GBM bursts \citep{Preece2016}. 
Here we compare the $E_p$ histograms of 32 SN/GRBs and 12 KN/GRBs with measured X-ray afterglow in Figure \ref{fig:logep}, from which a K-S test returns the statistic $D=0.25$ less than the critical value of ${D_\alpha }({n_1},{n_2}) = 0.46$ and the p-value $p=0.57$ larger than the significance level $\alpha=0.05$\textbf{ as illustrated in Table \ref{tab:7}}, which indicates they are drawn from the same parent distribution as illustrated for the total samples in Figure \ref{fig2:alpha-Ep}. 

\section{Conclusions}
\label{sec: conclusion}

Through our comprehensive and comparative analysis of the SN- and KN-associated GRBs, we can draw the following conclusions.
\begin{itemize}
\item The redshifts and durations of SN/GRBs and KN/GRBs are differently distributed. In contrast, the KN/GRBs having shorter durations are mainly produced from more nearby universe. The intrinsic durations of two kinds of bursts are bimodally distributed.
\item The peak luminosity distributions of SN/GRBs and KN/GRBs can be well fitted by a triple power-law relation, respectively. We observationally define the SN/GRBs with the bolometric luminosity higher than $6.5\times{10^{51}}$erg s$^{-1}$ as  SLSN/GRBs.
\item Three spectral parameters ($\alpha$, $\beta$ and $E_p$) in Band function for two kinds of bursts are lognormally distributed and have consistent mean values with normal GRBs. The $E_p$ values of SN/GRBs span a wider range than the KN/GRBs.
 \item Very importantly, we find that the anti-correlations between the spectral lag and the isotropic luminosity do exist for both the SN/GRBs and the KN/GRBs, of which these special bursts were claimed to be outliers away from the luminosity-lag relation of normal GRBs proposed by \cite{2000ApJ...534..248N}. Furthermore, the different types of GRBs are somewhat overlapped although the power law indexes of three luminosity-lag relations differ obviously.
 \item Both SN/GRBs and KN/GRBs exhibit their own Amati relations that are roughly consistent with those of long and short GRBs, respectively.
\item The X-ray light curves of either SN/GRBs or KN/GRBs can be artificially classified into three subclasses in terms of their temporal profiles. It is excitingly found SN/GRBs and KN/GRBs with measured X-ray afterglows have the identical $E_p$ distributions, resembling the total SN/KN GRB samples.
\item Using 7 KN/GRBs and 21 SN/GRBs with well-modelled X-ray plateaus, we discover the X-ray luminosity and plateau relation of $L_{X,KN}\propto T_a^{-1.12\pm0.17}$ for the first time and confirm the famous relation of $L_{X,SN}\propto T_a^{-1.08\pm0.22}$ for SN/GRBs. The uniform power-law indices imply that both types of GRBs could be produced by the same physical mechanism.
\end{itemize}

\section*{Data Availability}

The data underlying this article are available in the article (Zhang et al. 2018, PASP, 130, 054202), Gamma-ray Coordinates Network (GCN) at https://gcn.gsfc.nasa.gov/, https://www.mpe.mpg.de/~jcg/grbgen.html and The official \textit{Swift} website https://swift.gsfc.nasa.gov/results/batgrbcat/.
\section*{Acknowledgements}
This study was partly supported by the National Natural Science Foundation of China (Grant Nos. U2031118) and the science research grants from the China Manned Space Project with NO. CMS-CSST-2021-B11.



\bibliographystyle{mnras}
\bibliography{example} 

\appendix
\setlength{\tabcolsep}{0.5mm}{
\renewcommand\arraystretch{0.7}
\begin{center}
\onecolumn
\begin{longtable}{lccccccccccc}
\caption{53 SN/GRBs Included in Our Sample}
\label{tab:1} \\
\hline%
GRB & Instrument & ${T_{90}}$                         & z              & p                          & $S\gamma $                        &$\alpha$    & $\beta$     & $\textit{E}_{\rm p}$           & ${E_{\min}}$-${E_{\max}}$ & ${\Gamma _X}$ & ref$^{a,d}$   \\
      &          & (s)                         &             & (ph/cm$^{2}$/s) &$ (\times {\rm{1}}{{\rm{0}}^{{\rm{- 7}}}}erg/c{m^2}/s)$ &      &     &  (keV)             & (keV)     &               &     \\
(1)      & (2)        & (3)                         & (4)            & (5)                        & (6)                              &(7)   & (8)   & (9)           &(10)      & (11)           & (12)  \\
\hline%
\endhead%
\hline%
\endfoot%
\hline%
\endlastfoot%
970228$^*$   & Konus      & 56                          & 0.695          &2.59E-6$\pm$1.60E-7$^b$           & 82.90$\pm$5.50                      &  -1.92 & -     & 111.48$\pm$52.16$^c$  & 40-700    & -              & 1    \\
980326   & WFC        & 9                           & 0.9            & 3.14E-7$\pm$3.20E-8$^b$            & 50.90$\pm$5.00                       &  -1.85 & -     &  97.25$\pm$50.78$^c$    & 40-700    & -              & 2     \\
980425   & WFC        & 31                          & 0.009          & 1.71E-7$\pm$2.70E-8$^b$              & 14.10$\pm$1.50                       &  -1.90 & -     & 67.89$\pm$36.25$^c$   & 40-700    & -              & 3     \\
990712$^*$    &BeppoSAX   & 30                          & 0.433         & 5.72E-6$\pm$4.80E-7$^b$           & 109.90$\pm$9.10                  &  -1.36 & -     & 120.64$\pm$60.05$^c$  & 40-700    & -              & 2    \\
991208$^*$    & Konus      & 60                          & 0.706          & 2.0E-5$\pm$2E-6$^b$                        & 80.90$\pm$8.09                      &  -1.10 & -2.20 & 190.00$\pm$109.19 & 15-2000   & -              & 2     \\
000911$^*$    & Konus      & 500                         & 1.059          & 2E-5$\pm$2E-6$^b$                          & 220.00$\pm$20.00                     & -0.84 & -2.20 & 986.00$\pm$100.00 & 15-8000   & -              & 4,5   \\
011121   & WFC        & 105                         & 0.362          & 6.59E-06$\pm$5.30E-07$^b$           & 983.00$\pm$77.00                    &  -1.22 & -     & 222.80$\pm$10.90$^c$   & 25-100    & -              &6,7   \\
020305$^*$    & HETE       & 247                         & 1.98           & 4.70E-7$\pm$4.70E-8$^b$        & 30.00$\pm$3.00                       &  -1.06 & -2.30 &245.10$\pm$99.90  & 25-100    & -              &8     \\
020405$^*$    & Konus      & 40                          & 0.69           & 5.0E-6$\pm$2E-7$^b$                      & 49.50$\pm$4.95                      & -1.10 & -1.87 & 364.00$\pm$90.00  & 15-2000   & -              & 9     \\
020903   & HETE       & 13                          & 0.25           & 2.8$\pm$0.7                          & 1.00$\pm$0.30                        &  -    & -2.60 &5.00$\pm$0.80     & 30-400    & -              & 11    \\
021211   & HETE       & 8                           & 1.01           & 30$\pm$2                              & 35.30$\pm$2.10                       &  -0.81 & -2.37 & 46.80$\pm$5.10    & 30-400    & -              & 12,13 \\
030329   & HETE       & 63                          & 0.17           & 451$\pm$25                           & 1630.00$\pm$13.00                    &  -1.26 & -2.28 & 68.00$\pm$2.00   & 30-400    & -              & 10   \\
030723   & HETE       & 31                          & 0.54           & 2.1$\pm$0.4                           & 3.20$\pm$0.80                        &  -     & -1.90 & 8.90$\pm$0.89     & 30-400    & -              &14    \\
031203   & INTERFRAL  & 30                          & 0.105          & 1.3$\pm$0.13                                    & 20$\pm$0.4                          & -1.63 & -     & -                 & 50-300    & -              & 15,16 \\
040924$^*$    & HETE       & 5                           & 0.859          & 3.30E-6$\pm$3.5E-07$^b$          & 27.30$\pm$1.20                       &  -1.03 & -     &41.10$\pm$2.70    & 20-500    & -              & 17    \\
041006   & HETE       & 25                          & 0.716          & 8.4$\pm$0.3                           & 4.41$\pm$0.44                       &  -1.30 & -     & 47.70$\pm$2.70    & 30-400    & -              & 18    \\
050416A  & Swift      & 3                           & 0.65           & 4.4$\pm$0.80                                    & 0.37$\pm$0.05                        & -0.82 & -     & 14.85$\pm$6.97    & 15-150    & 2.06183        &47    \\
050525A$^*$   & Swift      & 9                           & 0.606          & 48 $\pm$1                                  & 15.10$\pm$0.20                      & -1.03 & -     & 79.00$\pm$7.90    & 15-150    & 2.08914        & 19    \\
050824   & Swift      & 23                          & 0.83           & 0.5$\pm$0.2                                     & 0.26$\pm$0.06                       & -1.00 & -     & 80.00$\pm$0.80    & 15-150    & 1.94709        & 20    \\
060218   & Swift      & 128                         & 0.033          & 0.156597                                        & 6.80$\pm$0.40                            & -1.65 & -     & 25.30$\pm$2.52    & 15-150    & 3.52495        & 10    \\
060729   & Swift      & 115.3                       & 0.54           & 1.4$\pm$0.2                                     & 2.61$\pm$0.23                        & -1.60 & -     & 201.16$\pm$20.11  & 15-150    & 2.02393        & 21    \\
060904B  & Swift      & 171.5                       & 0.703          & 2.5$\pm$0.2                                     & 1.62$\pm$0.18                      & -1.23 & -     & 84.09$\pm$8.01    & 15-150    & 2.18877        &47    \\
071112C  & Swift      & 15                          & 0.823          & -                                               & 3.00$\pm$0.30                        & -     & -     & -             & -         & 1.67267        & 22    \\
080319B$^*$   & Swift      & 125                         & 0.937          & 24.8$\pm$0.5                               & 85.00$\pm$0.83                       & -0.96 & -     & 999.06$\pm$99.91  & 15-150    & 1.81426        & 23  \\
081007   & Swift      & 10                          & 0.5295         & 2.6$\pm$0.4                                     & 0.58$\pm$0.08                        & -1.52 & -     & 20.12$\pm$2.12    & 15-150    & 2.102          & 24,25 \\
090618$^*$    & Swift      & 113                         & 0.54           & 38.9$\pm$0.8                               & 109.00$\pm$1.21                     & -1.49 & -     & 399.81$\pm$3.99   & 15-150    & 1.82831        & 21    \\
091127   & Swift      & 7                           & 0.49           & 46.5$\pm$2.7                                    & 8.58$\pm$0.33                        & -1.80 & -2.13 & 46.37$\pm$4.64    & 15-150    & 1.80187        & 26    \\
100316D  & Swift      & 521.88                      & 0.014          & 0.1$\pm$0.0                                     & 3.40$\pm$0.34                      & -1.88 & -2.35 & 9.62$\pm$0.96     & 15-150    & 2.49431        & 27   \\
100418A  & Swift      & 7                           & 0.62           & 1.0$\pm$0.2                                    & 0.34$\pm$0.03                        & 2.16  & -2.06 & 187.32$\pm$18.73  & 15-150    & 2.26662        & 28,29 \\
101219B  & Swift      & 34                          & 0.552          & 3.16$\pm$0.84                                  & 3.99$\pm$0.51                       & 1.38  & -2.26 & 56.36$\pm$6.64    & 10-1000   & 1.85684        & 30   \\
101225A  & Swift      & 1088                        & 0.847          & 0.094                                          & 1.90$\pm$0.50                        & -0.48 & -     & 96.65$\pm$9.66    & 15-150    & 1.81612        & 31   \\
111228A  & Swift      & 99.842                      & 0.72           & 27.58$\pm$1.74                                 & 8.06$\pm$0.28                       & -1.58 & -2.44 & 26.51$\pm$1.25    & 10-1000   & 2.03824        & 22    \\
111209A  & Swift      & 810.97                      & 0.677          & 0.5$\pm$0.1                                    & 36.00$\pm$0.59                       & -1.45 & -     & 768.79$\pm$76.88  & 15-150    & 1.81212        & 32    \\
120422A  & Swift      & 5.35                        & 0.28           & 0.6$\pm$0.2                                    & 0.30$\pm$0.10                       & 1.19  & -     & 97.01$\pm$9.70    & 15-150    & 2.32199        & 33    \\
120714B  & Swift      & 159                         & 0.3984         & 0.4$\pm$0.1                                    & 1.15$\pm$0.16                       & -0.69 & -     & 71.24$\pm$7.12    & 15-150    & 1.94091        & 22   \\
120729A  & Swift      & 25.472                      & 0.8            & 2.9$\pm$0.2                                    & 2.51$\pm$0.12                       & -1.61 & -     & 25.95$\pm$20.00   & 15-150    & 1.8831         & 34    \\
130215A  & Swift      & 65.7                        & 0.597          & 2.5$\pm$0.7                                   & 5.29$\pm$0.48                       & -1.18 & -     & 101.57$\pm$10.16  & 15-150    & -              &34    \\
130427A$^*$   & Swift      & 138.24                      & 0.34           & 331$\pm$4.6                               & 372.95$\pm$3.57                     & -1.17 & -     & 557.47$\pm$55.74  & 15-150    & 1.78644        & 35    \\
130702A  & Fermi      & 58.881                      & 0.145          & 16.51$\pm$4.69                                  & 5.72$\pm$0.57                       & 1.82  & -2.49 & 10.43$\pm$1.13    & 10-1000   & 1.98886        & 36    \\
130831A  & Swift      & 32.5                        & 0.4791         & 13.6$\pm$0.6                                    & 6.50$\pm$0.17                       & -1.91 & -     & 319.86$\pm$31.98  & 15-150    & 1.79449        & 22    \\
141004A  & Ferimi     & 9.472                       & 0.57           & 12.02$\pm$1.02                                 & 0.67$\pm$0.03                       & -0.18 & -2.67 & 63.95$\pm$2.72    & 15-150    & 1.79496        & 48    \\
140606B  & Swift      & 22.78                       & 0.384          & 16.26$\pm$1.35                                 & 9.02$\pm$0.37                        & -1.24 & -2.19 & 554.60$\pm$165.30 & 10-1000   & 1.91703        & 48   \\
150518   & MAXI       &> 1000                        & 0.256         & -                                              & …                                   & -                                  & -
& -      & -          & 3.6223         &38   \\
150818A  & Swift      & 123.3                       & 0.282          & 2.4$\pm$0.3                                    & 4.10$\pm$0.27                       & -1.88 & -     & 100.00$\pm$10.00  & 15-150    & 1.9776         & 37    \\
161219B  & Swift      & 6.94                        & 0.1475         & 4.4$\pm$0.4                                    & 1.50$\pm$0.07                        & -1.29 & -     & 61.88$\pm$6.18    & 15-150    & 1.86204        &37    \\
171010A  & Fermi      & 107.3                       & 0.3285         & 137.30$\pm$4.41                                & 633.00$\pm$1.50                      & -1.11 & -2.22 & 154 .00$\pm$1.00  & 10-1000   & 2.00598        &39   \\
171205A  & Swift      & 189.4                       & 0.0368         & 1.0$\pm$0.3                                    & 3.60$\pm$0.29                        & -1.22 & -     & 295.34$\pm$29.53  & 15-150    & 1.66173        & 40    \\
180720B$^*$   & Swift      & 108.4                       & 0.654          & 125$\pm$1                                  & 86.00$\pm$1.18                       & -1.34 & -     & 631.00$\pm$10.00  & 10-1000   & 1.8461         &41    \\
180728A  & Swift      & 8.68                        & 0.117          & 231$\pm$1.2                                   & 29.60$\pm$0.47                       & -1.91 & -     & 88.34$\pm$8.83    & 10-1000   & 1.70111        & 42    \\
181201A$^*$   & Intergral  & 172                         & 0.45           & 2.88E-5$\pm$3.30e-6$^b$                    & 1990$\pm$8.10                       & -1.25 & -2.73 & 152.00$\pm$6.00   & 20-15000  & 1.74367        &43    \\
190114C$^*$   & Swift      & 361.5                       & 0.42           & 246.86$\pm$0.86                          & 3990$\pm$8.10                        & -1.43 & -     & 998.60$\pm$11.90  & 10-1000   & 1.80689        & 45    \\
190829A  & Swift      & 58.2                        & 0.0785         & 25.6                                           & 127$\pm$1.5                         & -1.37 & -     & 18.90$\pm$1.88    & 10-1000   & 2.0913         & 44    \\
200826A$^*$   & IPN        & 1.14                        & 0.748          & 64.3$\pm$2.1                             & 42.6$\pm$0.2                        & -0.26 & -2.40 & 88.90$\pm$3.20    & 10-1000   & 1.54702        & 46   \\
\end{longtable}

\begin{tablenotes}
\item[a]{$a$References for the spectral parameters ($\textit{E}_{\rm p}$, $\alpha $, $\beta $, P, $S\gamma$ ${E_{\min}}$, ${E_{\max}}$, z, and $\Gamma_x$): \url{ https://www.mpe.mpg.de/~jcg/grbgen.html};\url{https://swift.gsfc.nasa.gov/results/batgrbcat/};\url{http://butler.lab.asu.edu/swift/bat_spec_table.html};The \textit{Beppo}SAX GRB parameters information come from \citep{2009ApJS..180..192F};The \textit{HETE-2} parameters information is obtained from \citep{2005ApJ...629..311S,2004ApJ...617.1251V,2008A&A...491..157P};The \textit{Swift} and \text{Fermi} GRBs information come from \citep{2007ApJ...660...16S,2015ApJS..218...13Y} and \citep{2018PASP..130e4202Z}}
\item[b]{$b$ These peak fluxes are in units of erg cm$^{-2}$  s$^{-1}$.}
\item[*]{$^* $ They are superluminous SN-associated GRBs.}
\item[C]{$^c $ The peak energy is simulated by the empirical formula of \citet{2020ApJ...902...40Z}.}
\item[d]{$d$Some literatures provide GRB and supernova association samples:(1)\citet{2000ApJ...536..185G};(2)\citet{2001BSAO...51...38S}; (3)\citet{2000ApJ...536..185G}; (4)\citet{2003AIPC..662..393C} (5)\citet{2001A&A...378..996L}; (6)\citet{2002ApJ...572L..45B};(7)\citet{2002AAS...200.2505G};(8)\citet{2005A&A...437..411G}; (9)\citet{2003ApJ...589..838P};(10)\citet{2017IAUS..331...45M};(11)\citet{2006ApJ...643..284B};(12)\citet{2003A&A...406L..33D};(13)\citet{2002GCN..1781....1F};
 (14)\citet{2004ApJ...612L.105T};(15)\citet{2004MSAIS...5..160C};(16)\citet{2004A&A...419L..21T};(17)\citet{2006ApJ...636..391S};
    (18)\citet{2006IAUC.8696....1D} ;(19)\citet{2006IAUC.8696....1D};(20)\citet{2007A&A...466..839S};(21)\citet{2011MNRAS.413..669C};(22)\citet{2019A&A...622A.138K};(23)\citet{2010ApJ...725..625T};(24)\citet{2008GCN..8662....1S};(25) \citet{2008CBET.1602....1D};(26)\citet{2010ApJ...718L.150C};(27)\citet{2010arXiv1004.2262C}; (28)\citet{2010GCN.10661....1H};(29)\citet{2012PASJ...64..115N}; (30)\citet{2011ApJ...735L..24S};(31)\citet{2013ApJ...778...67N};(32) \citet{2015Natur.523..189G};(33)\citet{2012A&A...547A..82M}; (34)\citet{2014A&A...568A..19C};
    (35)\citet{2013ApJ...776...98X};(36)\citet{2013GCN.14998....1C}; (37)\citet{2017IAUS..331...39D}; (38)\citet{2016AAS...22734003A};(39)\citet{2019MNRAS.490.5366M}; (40) \citep{2017ATel11038....1D};(41)\citep{2021APS..MARU71238M};(42) \citep{2018GCN.23142....1I}; (43)\citet{2020AstL...46..783B}; (44)\citet{2019GCN.25651....1B}; (45)\citet{2022A&A...659A..39M}; (46)\citet{2022ApJ...932....1R};(47)\citet{2012ApJ...746..156C}; (48)\citet{2020ApJ...904...97D} (49)\citet{2020A&A...641L..10M};}
\end{tablenotes}
\end{center}

\setlength{\tabcolsep}{1mm}{
\renewcommand\arraystretch{1.1}
\begin{center}
\onecolumn

\begin{longtable}{lccccccccccc}
\caption{15 KN/GRBs Included in Our Sample}
\label{tab:2} \\
\hline%
GRB & Instrument & ${T_{90}}$                         & z              & p                          & \textbf{$S\gamma $}                        &$\alpha$    & $\beta$     & $\textit{E}_{\rm p}$           & ${E_{\min}}$-${E_{\max}}$ & ${\Gamma _X}$ &ref$^{a,d}$  \\
      &          & (s)                         &             & (ph/cm$^{2}$/s) &$ (\times {\rm{1}}{{\rm{0}}^{{\rm{- 7}}}}erg/c{m^2}/s)$ &      &     &  (keV)             & (keV)     &               &     \\
(1)      & (2)        & (3)                         & (4)            & (5)                        & (6)                              &(7)   & (8)   & (9)           &(10)      & (11)           & (12)  \\
\hline%
\endhead%
\hline%
\endfoot%
\hline%
\endlastfoot%
050709   & HETE       & 0.07  & 0.1606   &34.10$\pm$2.70 & 10.00$\pm$1.00  & -0.82  & -     & 86.50$\pm$18.00   & 30-400    & -    &  1    \\
050724   & swift      & 96    & 0.257    & 3.35$\pm$0.31  &8.90$\pm$2.17    & -2.02  & -     & 78.91$\pm$8.00    & 15-150    & 1.80056          & 2    \\
060614A  & swift      & 102   & 0.125    & 11.60$\pm$0.70 & 188.29$\pm$14.56 & -2.037 & -     & 98.52$\pm$9.80    & 15-150    & 1.89544          &3  \\
061201   & swift      & 0.76  & 0.111    & 3.55$\pm$0.40  & 53.20$\pm$7.00   & -0.36  & -     & 873.00$\pm$458.00 & 20-3000   & 1.65934          & 12   \\
070809   & swift      & 1.3   & 0.2187   & 0.97$\pm$0.12  &0.73$\pm$0.09    & -1.33  & -     & 145.50$\pm$70.00  & 15-150    & 1.37246          & 4    \\
070714B  & swift      & 64    & 0.92     & 1.20$\pm$0.20  & 7.23$\pm$1.05    & -1.431 & -     & 111.13$\pm$11.11  & 15-150    & 2.07429          & 11   \\
080503$^b$  & swift      & 170   & 0.125 & 0.90$\pm$0.10  & 20.00$\pm$1.70   & -1.942 & -     & 73.22$\pm$7.32    & 15-150    & 2.5274           & 5    \\
100625A  & swift      & 0.33  & 0.452    & 11.00$\pm$0.30 &2.46$\pm$0.19    & -0.87  & -     & 319.30$\pm$100.00 & 15-150    & 2.40449          & 12   \\
111117A  & swift      & 0.47  & 2.211    & 1.35$\pm$0.20  & 6.70$\pm$0.20    & 0.051  & -     & 233.63$\pm$23.36  & 15-150    & 2.19593          &6    \\
130603B  & swift      & 0.18  & 0.3564   & 1.30$\pm$0.20  &6.24$\pm$0.28    & -1.58  & -     & 997.50$\pm$16.52  & 15-150    & 2.20112          & 7    \\
140903A  & swift      & 0.3   & 0.351    & 6.40$\pm$0.30  & 1.41$\pm$0.10    & -0.75  & -     & 44.17$\pm$16.20   & 15-150    & 1.63218          & 12   \\
150101B  & swift      & 0.018 & 0.093    & 2.50$\pm$0.20  & 1.09$\pm$0.14    & -1.361 & -1.94 & 96.55$\pm$9.66    & 15-150    & 2.30384          & 13   \\
160821B  & swift      & 0.5   & 0.16     & 9.16$\pm$1.19  &1.07$\pm$0.16    & -0.12  & -     & 46.32$\pm$19.00   & 10-1000   & 2.39723          & 8   \\
170817A  & fermi      & 2     & 0.009783 & 3.73$\pm$0.93  & 2.79$\pm$0.17    & 0.15   & -8.94 & 214.70$\pm$56.6   & 10-1000   & -                & 9    \\
200522A  & swift      & 0.62  & 0.554    & 1.50$\pm$0.20  & 1.04$\pm$0.14    & -1.452 & -     & 77.76$\pm$7.78    & 15-150    & 1.38048          &10    \\
\end{longtable}

\begin{tablenotes}
\item[a]{a References for the spectral parameters ($\textit{E}_{\rm p}$, $\alpha $, $\beta $, P, $S\gamma $, ${E_{\min}}$, ${E_{\max}}$, z, and $\Gamma_x$): \url{ https://www.mpe.mpg.de/~jcg/grbgen.html}; \url{https://swift.gsfc.nasa.gov/results/batgrbcat/}; \url{http://butler.lab.asu.edu/swift/bat_spec_table.html};The \textit{HETE-2} GRB 050709 parameters information is obtained from \citep{2005Natur.437..855V}}
\item[b]{b The redshift of GRB 080503A has been assumed to be 0.125 since it is very similar to GRB 060614, of which both of them have very large  ${T_{90}}$ but are associated with KNe.}
\item[c]{d Some literatures provide GRB and kilonova association samples:(1)\citet{2016NatCo...712898J};(2)\citet{2018ApJ...855...67L};(3)\citet{2015ApJ...811L..22J};(4)\citet{2020NatAs...4...77J};(5)\citet{2016ChA&A..40..439W};(6)\citet{2018A&A...616A..48S};
    (7)\citet{2013Natur.500..547T};(8)\citet{2018ApJ...857..128J};(9)\citet{2017Natur.551...67P};(10)\citet{2021MNRAS.502.1279O};(11)\citet{2017ApJ...837...50G}
    (12)\citet{2020ApJ...904...97D};(13)\citet{2020MNRAS.493.3379R};}
\end{tablenotes}
\end{center}}

\begin{table*}
	\centering
	\caption{A gold sample of 30 \textit{Swift} SN/GRBs with well-measured $\gamma$-ray components above 3$\sigma$.}
\begin{threeparttable}
	\label{tab:3}
	\begin{tabular}{lcccccc} 
		\hline
GRB & Subclass & Precursor & Main peak & Pulsating type & EE & triggered\\
\hline
        050416A & B        & Y         & Y  & S                      & N  & N          \\
		050525A & C        & N         & Y  & M                      & N  & N         \\
		050824  & B        & Y         & Y  & S                      & N  & N          \\
		060218  & B        & Y         & N  & N                      & Y  & N         \\
		060729  & A        & Y         & Y  & M                      & Y  & N          \\
		060904B & B        & Y         & Y  & S                       & N  & N        \\
		080319B & C        & N         & Y  & M                      & N  & Y         \\
		081007  & B        & Y         & Y  & S                      & N  & N        \\
		090618  & B        & Y         & Y  & M                      & N  & Y          \\
		091127  & C        & N         & Y  & M                      & N  & Y         \\
		100316D & A        & Y         & Y  & S                      & Y  & N        \\
		100418A & A        & Y         & Y  & S                      & Y  & N          \\
		101219B & A        & Y         & Y  & M                      & Y  & N         \\
		101225A & A        & Y         & Y  & S                      & Y  & N         \\
		111209A & A        & Y         & Y  & S                      & Y  & N         \\
		111228A & A        & Y         & Y  & M                      & Y  & Y          \\
		120422A & B        & Y         & N  & N                      & Y  & N          \\
		120714B & C        & N         & Y  & S                      & N  & N        \\
		120729A & B        & N         & Y & S                      & Y  & Y           \\
		130215A & B        & N        & Y  & S                      & Y  & Y         \\
		130427A & B        & N         & Y  & S                      & Y  & N         \\
		130831A & A        & Y         & Y  & S                      & Y  & N        \\
		141004A & B        & Y         & Y  & S                      & N  & N          \\
		150818A & A        & Y         & Y  & S                      & Y  & N         \\
		161219B & A        & Y         & Y  & S                      & Y  & N        \\
		171205A & A        & Y         & Y  & S                      & Y  & Y        \\
		180720B & A        & Y         & Y  & M                      & Y  & N         \\
		180728A & B        & Y         & Y  & S                      & N  & Y         \\
		190114C & C        & N         & Y  & M                      & N  & Y          \\
		190829A & B        & Y         & Y  & S                      & N  & N          \\
\hline
	\end{tabular}
\begin{tablenotes}
\item[a]{1.GRB 071112C is not involved because of its absence of spectral data;  2. The Y and N symbols respectively indicate that the corresponding components are existent and inexistent. The main peaks are marked with M and S to show whether the light curves consist of single or multiple pulses; 3. To symbolize how many components are included in prompt $\gamma$ emissions, namely precursors, main peaks and EEs, we use the symbols A, B and C to denote 3, 2 and 1 segments, individually.}
\end{tablenotes}
\end{threeparttable}
\end{table*}

\begin{table*}
	\centering
	\caption{A gold sample of 13 \textit{Swift} KN/GRBs with well-measurements above 3$\sigma$.}
\begin{threeparttable}	
\label{tab:4}
	\begin{tabular}{lcccccc} 
		\hline
GRB & Subclass & Precursor & Main peak & Pulsating type & EE & triggered\\
\hline
050724A & B        & N         & Y  & S                      & Y  & Y         \\
		060614  & A       & Y        & Y  & M                      & Y  & Y        \\
		061201  & A        & Y       & Y  & M                      & Y  & N        \\
		070809  & C        & Y       & Y  & S                      & N  & Y       \\
		070714B & A        & Y         & Y  & S                      & Y  & N         \\
		080503A & C        & N         & Y  & M                      & N  & Y        \\
		100625A & B        & N         & Y  & M                      & Y  & Y        \\
		111117A & B        & N         & Y  & M                      & Y  & Y        \\
		130603B & A        & Y         & Y  & S          & Y  & Y       \\
		140903A & B        & Y         & Y  & S                      & N  & N        \\
		150101B & C        & N         & Y  & S                      & N  & Y        \\
		160821B & B        & N         & Y  & S                      & Y  & Y         \\
		200522A & C        & N         & Y  & S                      & N  & Y        \\
		\hline
\end{tabular}
\begin{tablenotes}
\item[a]{The symbols in this table are the same as in Table \ref{tab:3}.}
\end{tablenotes}
\end{threeparttable}
\end{table*}


\setlength{\tabcolsep}{0.8mm}{
\renewcommand\arraystretch{0.85}
\begin{center}
\onecolumn
\begin{table*}
\centering
\caption{Supplementary parameters of 53 SN/GRBs included in our sample(N=53).}
\label{tab:5}
\resizebox{\linewidth}{!}{
	\begin{tabular}{lccccccccccccc} 
		\hline
GRB &$k_{c,\gamma}$&$k_{c,x}$ & $\tau (\rm PRE)$ &$\tau (\rm MP)$ &$\tau (\rm EE)$ &${P_{bolo}}$&$\Delta {P_{bolo}}$&${S_{bolo}}$& $\Delta {S_{bolo}}$&${L_{{\rm p}}}$  & ${\rm{Lp \pm}}$&${E_{iso}}$ &${E_{iso}}{\rm{\pm}}$\\
    && & $ s$ & $s$ &$s$ &${10^{- 7}}{\rm{erg}}/c{m^2}/s$&${10^{- 7}}{\rm{erg}}/c{m^2}/s$&${10^{- 6}}{\rm{erg}}/c{m^2}/s$&${10^{- 6}}{\rm{erg}}/c{m^2}/s$&${10^{51}}erg/s$  & ${10^{51}}erg/s$               &${10^{52}}erg$   &${10^{52}}erg$   \\
  (1) &(2) &(3) &(4) &(5) &(6)&(7)&(8)&(9)&(10)&(11)&(12)&(13)&(14)\\
	\hline
   970228   & 2.57 & -      & -           & -            & -           &3.53E+01     & 2.18E+00& 1.13E+01     & 7.49E-01 & 7.54E+00   & 4.66E-01 & 1.42E+00  & 9.44E-02   \\
  980326   & 2.37   & -      & -           & -            & -           & 8.16E+00     & 8.31E-01 & 1.32E+01     &1.30E+00 &  3.29E+00   & 3.35E-01 & 2.81E+00  & 2.76E-01   \\
  980425   & 2.39   & -      & -           &4.50$^{a}$        & -           & 3.97E+00     & 6.27E-01 & 3.27E+00     &  3.48E-01 & 7.16E-05   & 1.13E-05 & 5.85E-05  & 6.23E-06   \\
  990712   & 1.59   & -      & -           & -            & -           & 1.40E+02     & 1.18E+01  & 2.69E+01     &2.23E+00 & 9.56E+00   & 8.02E-01 &1.28E+00  & 1.06E-01   \\
  991208   & 1.42   & -      & -           & -            & -           & 2.19E+02     & 2.19E+01  & 8.88E+01    & 8.88E+00 & 4.87E+01   &4.87E+00&1.16E+01  &1.16E+00   \\
  000911   & 1.04   & -      & -           & -            & -           & 2.01E+02     &  2.01E+01  & 2.21E+02     & 2.01E+01 & 1.21E+02   & 1.21E+01 &6.47E+01  & 5.88E+00  \\
  011121   & 1.37   & -      & -           & -            & -           & 9.65E+01     &  7.76E+00  & 1.44E+02     & 1.13E+01 & 4.32E+00   & 3.47E-01 & 4.73E+00  & 3.70E-01  \\
  020305   & 1.28  & -      & -           & 0.44 ± 0.04  & -           & 4.78E+00      &4.78E-01  &2.05E-03    & 6.15E-04 & 1.35E+01   & 1.35E+00 & 1.94E-03 & 5.83E-04  \\
  020405   & 1.31   & -      & -           & -            & -           & 5.29E+01     &  2.12E+00  & 5.24E+01     & 5.24E+00 & 1.11E+01  &4.44E-01 & 6.51E+00  & 6.51E-01   \\
  020903   & 1.37   & -      & -           & -            & -           & 2.07E-01     & 5.17E-02  & 1.27E-01     & 3.83E-02 & 3.93E-03  & 9.83E-04 & 1.94E-03  & 5.83E-04   \\
  021211   & 4.27   & -      & -           & 0.59 ± 0.02 & -           & 7.11E+00     & 4.74E-01  & 3.64E+00     & 2.16E-01 & 3.81E+00   & 2.54E-01 &9.70E-01  &5.77E-02   \\
  030329   & 2.78  & -      & -           & -            & -           & 1.05E+02     & 5.85E+00  & 1.68E+02     & 1.34E+00 &  8.46E-01   &4.69E-02 & 1.15E+00  & 9.17E-03   \\
  030723   & 2.62   & -      & -           & -            & -           & 1.71E-01     & 3.26E-02  & 3.63E-01     & 9.08E-02 &  1.98E-02   &3.77E-03 & 2.73E-02  &6.82E-03   \\
  031203   & 1.64   & -      & -           & 0.22 ± 0.02  & -           & 1.73E+01     & 1.73E+00  & 1.41E+01     & 2.82E+00 & 4.89E-02   & 4.89E-03 & 3.59E-02  & 7.18E-03   \\
  040924   & 2.62   & -      & -           & 0.44 ± 0.03  & -           & 5.34E+01     &  5.66E+00  & 4.42E+00     & 1.94E-01 & 1.92E+01   & 2.04E+00 & 8.54E-01 & 3.75E-02   \\
  041006   & 2.43   & -      & -           &-0.03 ± 0.01 & -           & 1.70E+01     &7.07E-01  & 8.88E-01    & 8.86E-02 & 3.90E+00  & 1.62E-01 & 1.19E-01  &1.19E-02   \\
  050416A  & 16.78 & 1.70   & -           & -            & -           & 4.54E+00     &8.25E-01  & 9.75E-01     & 1.32E-01 & 8.24E-01   & 1.50E-01 & 1.07E-01  &1.45E-02   \\
  050525A  & 1.66   &1.68   & -           & 0.56 ± 0.14  & -           & 5.02E+01     & 1.05E+00  & 2.24E+01    &  2.96E-01 &7.68E+00   & 1.60E-01 &2.13E+00  & 2.82E-02   \\
  050824   & 1.64   &1.77   & -           & -            & -           & 5.23E-01     &2.09E-01  & 3.76E-01     & 8.23E-02 & 1.73E-01   & 6.91E-02 & 6.79E-02  &1.49E-02   \\
  060218   & 3.75   & 1.09   & -           & 81.00±12.00$^{a}$  & -           & 6.53E-01     &6.53E-02  & 1.42E+01     & 8.36E-01 &1.64E-04   & 1.64E-05 & 3.46E-03  & 2.04E-04   \\
  060729   & 1.74   &1.56   & -           &1.02 ± 0.19  & 1.73± 0.23  & 2.45E+00     & 3.50E-01  & 6.57E+00    & 5.89E-01 & 2.84E-01   &4.05E-02 &4.93E-01  & 4.43E-02  \\
  060904B  &1.76   & 1.88   & -           &1.09 ± 0.13  & -           & 2.83E+00     & 2.27E-01  & 2.64E+00     & 2.90E-01 &6.23E-01   & 4.98E-02 &3.41E-01  &3.74E-02  \\
  071112C  & -      & 1.50  & -           & -            & -           & -            & -        & -            & -        & -          & -        & -         & -          \\
  080319B  &2.15   &1.71   & -           &2.30 ± 0.06  & -           & 1.92E+02     &3.87E+00  & 7.08E+02     & 6.89E+00 & 8.56E+01  &1.73E+00 &1.63E+02  &1.58E+00   \\
  081007   &5.33   &1.60   & -           & 0.33± 0.03   & -           & 2.93E+00     & 4.50E-01  & 1.31E+00     & 1.86E-01 &3.23E-01   &4.97E-02 & 9.42E-02  &1.34E-02   \\
  090618   & 1.70   & 1.43   & 4.62 ± 0.04 &1.26 ± 0.11  &2.31 ± 0.13 &9.13E+01      & 1.88E+00 & 3.45E+02    &3.83E+00 &1.06E+01   &2.17E-01 &2.59E+01  &2.88E-01   \\
  091127   & 4.30   & 1.38   & -           &0.09 ± 0.05  & -           & 6.61E+01     &3.84E+00  & 1.99E+01     & 7.77E-01 &6.05E+00  & 3.51E-01 &1.23E+00  &4.78E-02   \\
  100316D  & 21.67  & 1.02   & -           & -            & -           & 2.64E+00     & -        &1.66E+02      & 1.66E+01 & 1.16E-04   & -        &7.19E-03  &7.19E-04   \\
  100418A  & 1.02   & 1.84  & -           & -0.71±0.04   & -           & 3.75E+00     &7.50E-01  & 8.64E-01     &  8.64E-02 & 6.06E-01  & 1.21E-01 & 8.63E-02  & 8.63E-03   \\
  101219B  & 2.20   &1.46   & -           & -            & -           & 2.15E+00     &5.73E-01  & 4.04E+00     & 5.20E-01 &  2.63E-01  &6.99E-02 & 3.17E-01  &4.09E-02   \\
  101225A  &1.34   &1.65   & -           & -            & -           &1.06E-01     & -        &3.36E+00      & 8.83E-01 &  3.67E-02   & -        &6.32E-01  & 1.66E-01   \\
  111228A  & 10.51  &1.76   & -           &0.06 ± 0.11  &0.22 ± 0.12 &1.10E+01      & 6.97E-01 & 1.35E+01     & 4.70E-01 & 2.57E+00   &1.62E-01 & 1.83E+00  & 6.37E-02   \\
  111209A  & 1.98   &1.52   & -           & -            & -           & 1.67E+00     & 3.34E-01  & 1.57E+02     & 2.56E+00 & 3.34E-01   & 6.69E-02 & 1.88E+01  &3.05E-01   \\
  120422A  &1.13   & 1.39  & -           & 1.61 ± 0.13  & -           & 7.26E-01    &2.42E-01  & 3.60E-01     & 1.18E-01 & 1.79E-02   &5.97E-03 &6.93E-03  &2.28E-03   \\
  120714B  & 1.62   & -      & -           & -            & -           & 3.74E-01     & 9.36E-02 &1.51E+00      & 2.13E-01 &  2.10E-02   &5.24E-03 &6.05E-02  & 8.54E-03   \\
  120729A  &3.96   &1.68   & -           & 0.83 ± 0.06  & -           & 3.41E+00     &2.35E-01  & 1.09E+01    & 2.64E-01 &1.03E+00   &7.09E-02 & 1.83E+00  &4.43E-02   \\
  130215A  &1.56   & -      & -           & 0.07± 0.11   & -           & 3.03E+00     & 8.47E-01 & 8.82E+00      & 7.94E-01 & 4.46E-01   &1.25E-01 & 8.15E-01 & 7.33E-02   \\
  130427A  & 1.97   & -      & -           &1.08± 0.12  & -           & 1.20E+03    & 1.66E+01 & 1.60E+03      & 1.53E+01 & 4.63E+01   &6.43E-01 & 4.60E+01  & 4.41E-01   \\
  130702A  &1.61   & 1.14   & -           & -            & -           & 7.94E+00     & 2.25E+00 & 1.23E+01      & 1.23E+00 & 4.49E-02   &1.28E-02 &6.09E-02  &6.09E-03   \\
  130831A  &2.68   & 1.36  & -           & 0.48± 0.06   & -           & 3.05E+01     & 1.34E+00  & 2.30E+01     &  5.95E-01 & 2.65E+00   & 1.17E-01 &1.35E+00  &3.49E-02   \\
  141004A  & 1.97   & 1.43   & -           &0.28± 0.05   & -           & 1.01E+01     & 1.01E+00  & 7.89E-01     &  3.80E-02 & 1.33E+00   &1.33E-01 &6.63E-02  &3.19E-03   \\
  140606B  & 1.20   &1.35   & -           & -            & -           & 3.49E+01     &2.90E+00  & 9.62E+00     & 4.67E-01 & 1.79E+00   &1.49E-01 & 3.57E-01 & 1.73E-02   \\
  150518   & -     &1.82  & -           & -            & -           & -            & -        & -            & -        & -          & -        & -         & -          \\
  150818A  & 2.35   & 1.27   & -           &0.57 ± 0.10  & -           & 4.22E+00     & 5.27E-01 & 1.14E+01     & 7.40E-01 & 1.06E-01   & 1.32E-02 &2.23E-01  &1.45E-02   \\
  161219B  &2.03   & 1.13   & -           &0.13 ± 0.09  & -           & 4.57E+00     & 4.16E-01 & 2.40E+00     & 1.16E-01 &2.69E-02  &2.44E-03&1.23E-02  &5.95E-04   \\
  171010A  & 1.54   & 1.33   & -           & -            & -           & 1.49E+02     & 4.77E+00 & 6.84E+02     &1.62E+00 & 5.29E+00   & 1.70E-01 & 1.84E+01  &4.35E-02   \\
  171205A  & 1.38    & 1.02   & -           &15.54 ± 2.09 & -           & 2.23E+00     &6.68E-01 & 9.95E+00     & 8.11E-01 & 7.00E-04   &2.10E-04 &3.02E-03  &2.46E-04   \\
  180720B  & 1.78   & 1.53   & -           &0.70 ± 0.13  & 0.31 ± 0.05 & 2.66E+02     &2.13E+00 & 1.16E+02     & 1.59E+00 &4.90E+01   & 3.92E-01 & 1.29E+01  &1.78E-01   \\
  180728A  & 2.41   &1.08  & -           &-0.07 ± 0.03 & -           & 2.54E+02     & 1.32E+00 &4.55E+01    & 7.16E-01 & 9.04E-01   &4.70E-03 & 1.45E-01  & 2.28E-03   \\
  181201A  & 1.51   & 1.32   & -           & -            & -           & 3.61E+02     & 4.13E+01 &2.50E+02    & 1.00E+00 & 2.70E+01   &3.09E+00 & 1.29E+01  &5.17E-02   \\
  190114C  & 2.18  & 1.33   & -           & 0.37 ± 0.15  & 0.34 ± 0.13 & 6.10E+02     & 2.12E+00 & 1.33E+02     & 1.52E+00 &3.87E+01   &1.35E-01 & 5.93E+00  & 6.79E-02   \\
  190829A  & 6.26   & 1.09  & -           & -0.57 ± 0.10 & -           &1.61E+01     & -        &1.11E+01     &2.03E-01 & 2.44E-02  & -        & 1.56E-02  &2.87E-04   \\
  200826A  & 1.56   &1.36  & -           & -            & -           & 4.38E+01     &1.42E+00 &4.46E+00     & 2.10E-02 & 1.12E+01   & 3.64E-01 & 6.53E-01  & 3.07E-03\\
\hline
\end{tabular}}
\begin{tablenotes}
\item[a]{$^a$ The spectral lag of GRB 980425 and GRB 060218 are taken from \citep{2000ApJ...534..248N,2006ApJ...653L..81L} respectively.}
\end{tablenotes}
\end{table*}
\end{center}

\begin{table*}
	\centering
	\caption{Supplementary parameters of 15 KN/GRBs included in our sample(N=15).}
	\label{tab:6}
	\begin{tabular}{lcclccccccccc} 
		\hline
GRB &$k_{c,\gamma}$&$k_{c,x}$  &$\tau (\rm MP)$ &$\tau (\rm EE)$ &${P_{bolo}}$&$\Delta {P_{bolo}}$&${S_{bolo}}$& $\Delta {S_{bolo}}$&${L_{{\rm p}}}$  & ${\rm{Lp \pm}}$&${E_{iso}}$ &${E_{iso}}{\rm{\pm}}$\\
    && & $ s$  &$s$ &${10^{- 7}}{\rm{erg}}/c{m^2}/s$&${10^{- 7}}{\rm{erg}}/c{m^2}/s$&${10^{- 6}}{\rm{erg}}/c{m^2}/s$&${10^{- 6}}{\rm{erg}}/c{m^2}/s$&${10^{51}}erg/s$  & ${10^{51}}erg/s$               &${10^{52}}erg$   &${10^{52}}erg$   \\
  (1) &(2) &(3) &(4) &(5) &(6)&(7)&(8)&(9)&(10)&(11)&(12)&(13)\\
	\hline
050709   & 1.35  & -       &  0.03 ± 0.02      &  -               &  5.53E+01    &  4.38E+00   &   4.10E-01    &  5.14E-02    &  3.91E-01    &  3.10E-02    &  2.50E-03    &  3.13E-04 \\
050724   & 5.37 & 1.20  &  0.01 ± 0.00      &  -               &  1.10E+01    & 1.02E+00    & 4.78E+00     &  1.16E+00   & 2.24E-01     & 2.07E-02     & 7.70E-02     &  1.88E-02 \\
060614A & 7.66  & 1.11  &  0.02± 0.00      &   0.08±0.03 &  5.43E+01    & 3.27E+00    & 1.09E+02     & 8.45E+00    & 2.22E-01     &  1.34E-02   & 3.98E-01     & 3.08E-02  \\
061201   & 1.02  & 1.07  &  0.01 ± 0.01     &  -                &  2.24E+01   & 2.53E+00    & 5.41E+00     & 7.12E-01     & 7.12E-02      & 8.02E-03    & 1.55E-02     &  2.03E-03 \\
070809   & 1.99  & 1.08  & -                    &  -                & 1.42E+00    & 1.76E-01     & 1.46E-01      & 1.81E-02     &  2.00E-02     & 2.47E-03     & 1.69E-03     &  2.09E-04 \\
070714B & 1.94  & 2.02  &  0.02 ± 0.00   &  -                 & 1.62E+00   & 2.70E-01     & 1.40E+00    & 2.04E-01     & 6.90E-01      & 1.15E-01     & 3.10E-01     &  4.52E-02 \\
080503   & 2.97  &1.20   &  0.04 ± 0.02   &  -                   & 1.66E+00   & 1.84E-01      & 5.94E+00     & 5.04E-01    & 6.79E-03      & 7.55E-04     & 2.17E-02    &  1.84E-03 \\
100625A & 3.14 & 1.69  & 0.03± 0.00    &  -                 & 3.15E+01   &  8.58E-01     & 9.71E-01     &   7.31E-02  & 2.38E+00        & 6.48E-02  &  5.05E-02    &  3.80E-03 \\
111117A &  2.80 & 4.04  &  0.002 ± 0.01   & -                 &  4.28E+00  &  6.34E-01    &  3.57E+00    & 1.04E-01   &  1.58E+01    &  2.34E+00   &  4.11E+00   &  1.20E-01 \\
130603B &  4.47 & 1.44  &  0.01 ± 0.00   &  -                 &  4.23E+00  &  6.50E-01   &  2.38E+00    &  1.06E-01   &  1.82E-01       &  2.80E-02   &  7.57E-02  &  3.37E-03 \\
140903A &  1.35 & 1.21  &  0.04 ± 0.00   &  -                 &  5.12E+00  &  2.40E-01    &  1.90E-01    &  1.38E-02   &  2.13E-01      &  9.98E-03     &  5.86E-03    &  4.26E-04 \\
150101B &  1.74 & 1.12  &  0.02 ± 0.00   &  -                   & 3.08E+00  &  2.47E-01   &  1.59E-01     &  2.04E-02   &  6.71E-03    &  5.37E-04    &  3.17E-04        &  4.07E-05 \\
160821B &  1.08 & 1.23  &  0.03 ± 0.00  &  -                & 5.25E+00  &  6.82E-01     &  1.32E-01    &  1.94E-02   &  3.69E-02      &  4.79E-03    &  8.00E-04       &  1.18E-04 \\
170817A &  1.00 & -     &  0.15$^a$             &  -             &  7.32E+00    &  1.83E+00   &  2.80E-01    &  1.75E-02   &  1.56E-04    &  3.90E-05   &   5.92E-06      &  3.69E-07 \\
200522A &  1.81 & 1.18   &  0.02 ± 0.00     &  -               &  1.81E+00  &  2.41E-01   &  1.88E-01    &  2.61E-02   &  2.22E-01     &  2.97E-02    &  1.49E-02        &  2.07E-03 \\
\hline
\end{tabular}
\begin{tablenotes}
\item[a]{$^a$ The spectral lag of GRB 170817A is taken from \cite{2017ApJ...848L..14G}.}
\end{tablenotes}
\end{table*}

\begin{table*}
	\centering
	\caption{The K-S test results of our sample with a significance level of $\alpha=0.05$.}
	\label{tab:7}

	\begin{tabular}{lccccccc} 

		\hline

Parameters  & Samples &$m $&$n $&$D$& ${D_{\alpha}(m,n)}$&$P$-value &Judgement\\
\hline
 z            & SN/GRB-KN/GRB    & 53 & 15 & 0.43 & 0.40 & 1.80E-02 & Rejected \\
 \hline
$\log{T_{90}}$     & SN/GRB-KN/GRB    & 50 & 15 & 0.71 & 0.40 & 3.14E-06 & Rejected \\
 \hline
$\log {T_{90/(1 + z)}}$ & SN/GRB-KN/GRB    & 50 & 15 & 0.69 & 0.40 & 7.10E-06 & Rejected \\
 \hline
$k_\gamma$ & SN/GRB-KN/GRB    & 51 & 15 & 0.22 & 0.40 & 5.30E-01 & Accepted \\
 \hline
$\log L$         & SN/GRB-KN/GRB    & 51 & 15 & 0.44 & 0.40 & 2.00E-02 & Rejected \\
             & SLSN/GRB-SN/GRB & 15 & 51 & 0.71 & 0.40 & 3.78E-06 & Rejected \\
             & SLSN/GRB-KN/GRB & 15 & 15 & 0.93 & 0.50 & 3.86E-07 & Rejected \\
              \hline
$\alpha$        & SN/GRB-KN/GRB    & 49 & 15 & 0.26 & 0.40 & 3.40E-01 & Accepted \\
             & SLSN/GRB-SN/GRB & 16 & 49 & 0.24 & 0.39 & 3.80E-01 & Accepted \\
             & SLSN/GRB-KN/GRB & 16 & 15 & 0.34 & 0.49 & 2.70E-01 & Accepted \\
              \hline
$\beta$         & SLSN/GRB-SN/GRB & 6  & 19 & 0.13 & 0.64 & 9.90E-01 & Accepted \\
 \hline
$\log E_p$       & SN/GRB-KN/GRB    & 50 & 15 & 0.27 & 0.40 & 3.30E-01 & Accepted \\
             & SLSN/GRB-SN/GRB$^\dag$ & 15 & 50 & 0.27  & 0.48 & 3.20E-01 & Accepted \\
             & SLSN/GRB-KN/GRB & 15 & 15 & 0.40  & 0.50 & 1.80E-01 & Accepted \\
 \hline
$\log E_{p,i}$     & SN/GRB-KN/GRB    & 52 & 15 & 0.21 & 0.40 & 5.90E-01 & Accepted \\
 \hline
$\log {E_{iso}}$    & SN/GRB-KN/GRB    & 51 & 15 & 0.51 & 0.40 & 3.00E-03 & Rejected\\
\hline
$\log E_p$$^\ddag$         & SN/GRB-KN/GRB    & 32 & 12 & 0.25 & 0.46 & 5.70E-01 & Accepted \\
\hline
\end{tabular}\\
Note: $^\dag$ A significance level of $\alpha=0.01$ has been used.\\
      $^\ddag$ The $E_p$ distributions of SN/KN GRBs are compared for Figure \ref{fig:logep}.\\
\end{table*}
\setlength{\tabcolsep}{1mm}{
\begin{table*}
	\centering
	\caption{Parameters of X-ray afterglows with good plateaus of 21 SN/GRBs and 7 KN/GRBs in our sample.}
	\label{tab:8}
\begin{threeparttable}
	\begin{tabular}{llccccccccc} 
	\hline
		&GRB & ${T_X}^{a} $ & ${{\Delta T_X}}^{a} $ &${F_X}^{b} $ &${\Delta F_X}^{b} $ &${a_1}$ &${a_2}$ & ${\Gamma}_X$ & ${L_X}^{c}$ & ${\Delta L_X}^{c}$\\
		\hline
       & 060729  & 5.72E+01 & 3.26E+00 & 1.85E+01 & 5.14E-01 & 0.10$\pm$0.01  & 1.38$\pm$0.08  & 2.02393 & 2.16E+01 & 6.01E-01 \\
       & 081007  & 3.00E+01 & 9.98E+00 & 3.09E+00 & 7.63E-01 & 0.05$\pm$0.29  & 1.49$\pm$0.42  & 2.102   & 3.56E+00 & 8.79E-01 \\
       & 090618  & 5.95E+00 & 1.31E+00 & 1.85E+02 & 3.44E+01 & 0.61$\pm$0.03  & 1.42$\pm$0.17  & 1.82831 & 1.99E+02 & 3.70E+01 \\
       & 111228A & 1.04E+01 & 3.77E+00 & 2.45E+01 & 5.88E+00 & 0.34$\pm$0.05  & 1.17$\pm$0.21  & 2.03824 & 5.75E+01 & 1.38E+01 \\
       & 050416A & 1.07E+00 & 5.33E-01 & 2.28E+01 & 6.24E+00 & 0.21$\pm$0.13  & 0.89$\pm$0.11  & 2.06183 & 4.32E+01 & 1.18E+01 \\
       & 060904b & 4.15E+00 & 1.88E+00 & 1.27E+01 & 6.57E+00 & 0.67$\pm$0.06  & 1.42$\pm$0.06  & 2.18877 & 3.09E+01 & 1.60E+01 \\
       & 100316D & 8.62E-01 & 7.27E-02 & 2.14E+03 & 7.75E+01 & 0.09$\pm$0.03  & 2.00$\pm$0.04  & 2.49431 & 9.48E-01 & 3.43E-02 \\
       & 100418A & 6.96E+01 & 1.45E+01 & 1.08E+00 & 1.07E-01 & -0.14$\pm$0.05 & 1.42$\pm$0.031 & 2.26662 & 2.01E+00 & 2.00E-01 \\
       & 101219B & 1.91E+01 & 3.30E+00 & 1.10E+00 & 7.66E-02 & -0.82$\pm$0.45 & 0.71$\pm$0.08  & 1.85684 & 1.26E+00 & 8.77E-02 \\
       & 101225A & 2.25E+01 & 4.53E-01 & 1.02E+02 & 3.89E+00 & 1.11$\pm$0.01  & 6.09$\pm$0.29  & 1.81612 & 3.16E+02 & 1.21E+01 \\
SN/GRBs & 120422A & 1.16E+01 & 6.75E+00 & 3.06E-01 & 3.44E-02 & -0.38$\pm$0.46 & 0.52$\pm$0.10  & 2.32199 & 8.37E-02 & 9.41E-03 \\
       & 120729A & 8.57E+00 & 9.20E-01 & 5.55E+00 & 9.78E-01 & 1.13$\pm$0.02  & 3.00$\pm$0.21  & 1.8831  & 1.56E+01 & 2.75E+00 \\
       & 130702A & 1.11E+02 & 1.23E+01 & 1.26E+01 & 1.29E+00 & -0.14$\pm$0.58 & 1.25$\pm$0.02  & 1.98886 & 7.12E-01 & 7.29E-02 \\
       & 130831A & 6.50E-01 & 3.29E-02 & 2.83E+02 & 1.22E+01 & -3.30$\pm$0.92 & 1.16$\pm$0.14  & 1.79449 & 2.27E+02 & 9.77E+00 \\
       & 141004A & 6.06E-01 & 3.17E-01 & 4.53E+01 & 1.58E+01 & 0.36$\pm$0.08  & 1.21$\pm$0.37  & 1.79496 & 5.45E+01 & 1.90E+01 \\
       & 161219B & 1.57E+00 & 1.39E-01 & 1.96E+02 & 7.01E+00 & -0.66$\pm$0.24 & 0.74$\pm$0.03  & 1.86204 & 1.13E+01 & 4.04E-01 \\
       & 171010A & 4.35E+02 & 1.98E+01 & 6.66E-01 & 4.94E-01 & 1.29$\pm$0.07  & 2.33$\pm$0.41  & 2.00598 & 3.45E-01 & 2.56E-01 \\
       & 171205A & 1.81E+02 & 7.21E+01 & 7.08E-01 & 1.28E-01 & 0.18$\pm$0.05  & 1.43$\pm$0.78  & 1.66173 & 2.20E-03 & 3.98E-04 \\
       & 180720B & 3.05E+00 & 6.59E-01 & 2.04E+03 & 3.40E+02 & 0.41$\pm$0.07  & 1.41$\pm$0.12  & 1.8461  & 3.48E+03 & 5.79E+02 \\
       & 180728A & 2.44E+01 & 4.78E+00 & 8.63E+01 & 1.87E+01 & 0.83$\pm$0.02  & 1.34$\pm$0.02  & 1.70111 & 2.97E+00 & 6.43E-01 \\
       & 190829A & 1.42E+00 & 2.93E-02 & 1.43E+03 & 1.58E+01 & -1.97$\pm$0.13 & 0.95$\pm$0.03  & 2.0913  & 2.16E+01 & 2.38E-01 \\
         \hline
       & 060614  & 3.89E+01 & 4.30E+00 & 8.09E+00 & 4.78E-01 & 0.00$\pm$0.04  & 1.65$\pm$0.23  & 1.89544 & 3.28E-01 & 1.94E-02 \\
       & 061201  & 7.32E-01 & 7.51E-01 & 8.14E+01 & 5.22E+01 & 0.21$\pm$0.40  & 1.31$\pm$0.69  & 1.65934 & 2.49E+00 & 1.60E+00 \\
       & 070809  & 7.19E+00 & 3.56E+00 & 3.00E+00 & 9.06E-01 & 0.03$\pm$0.15  & 1.44$\pm$0.70  & 1.37246 & 3.73E-01 & 1.13E-01 \\
       & 070714B & 1.36E+00 & 3.10E-01 & 2.67E+01 & 5.65E+00 & 0.31$\pm$0.17  & 2.00$\pm$0.01  & 2.07429 & 1.19E+02 & 2.52E+01 \\
KN/GRBs & 080503A & 2.23E-01 & 6.57E-03 & 1.14E+03 & 2.03E+02 & 1.71$\pm$0.07  & 5.38$\pm$0.17  & 2.5274  & 4.97E+01 & 8.86E+00 \\
       & 130603B & 2.78E+00 & 4.29E-01 & 3.19E+01 & 6.64E+00 & 0.35$\pm$0.07  & 1.66$\pm$0.06  & 2.20112 & 1.46E+01 & 3.04E+00 \\
       & 140903A & 6.17E+00 & 1.84E+00 & 1.28E+01 & 1.91E+00 & -0.12$\pm$0.12 & 1.06$\pm$0.18  & 1.63218 & 4.77E+00 & 7.11E-01\\
		\hline
	\end{tabular}
\begin{tablenotes}
\item[a]{The plateau times and errors are in units of ${\rm{1}}{{\rm{0}}^{\rm{3}}}\ {\rm{s}}$}.
\item[b]{The X-ray fluxes and errors at $T_X$ are in units of ${\rm{1}}{{\rm{0}}^{{\rm{-12}}}}{\rm{erg}}\ {{\rm{cm}}^{\rm{-2}}}\ {\rm{s^{-1}}}$}.
\item[c]{The X-ray luminosities and errors at $T_X$ are in units of ${10^{45}}\rm {erg}\ s^{-1}$}.
\end{tablenotes}
\end{threeparttable}
\end{table*}
}



\bsp	
\label{lastpage}
\end{document}